\documentclass[aps,prd,10pt,twocolumn,superscriptaddress,nofootinbib]{revtex4-2}

\usepackage{graphicx}
\usepackage{bm}
\usepackage{color}
\usepackage[colorlinks=true,pdfstartview=FitV,linkcolor=blue,citecolor=blue,urlcolor=blue]{hyperref}
\usepackage[dvipsnames]{xcolor}
\usepackage{amsmath, amsfonts}

\enlargethispage{\baselineskip}
\everymath{\displaystyle}

\begin{document}
\title{Testing Quintessence Axion Dark Energy with Recent Cosmological Results}

\author{Weikang Lin}
\email{weikanglin@ynu.edu.cn}
\affiliation{South-Western Institute for Astronomy Research, Key Laboratory of Survey Science of Yunnan Province, Yunnan University, Kunming, Yunnan 650500, People's Republic of China}
\author{Luca Visinelli}
\email{lvisinelli@unisa.it}
\affiliation{Dipartimento di Fisica ``E.R.\ Caianiello'', Universit\`a degli Studi di Salerno,\\ Via Giovanni Paolo II, 132 - 84084 Fisciano (SA), Italy}
\affiliation{Istituto Nazionale di Fisica Nucleare - Gruppo Collegato di Salerno - Sezione di Napoli,\\ Via Giovanni Paolo II, 132 - 84084 Fisciano (SA), Italy}
\author{Tsutomu T.\ Yanagida}
\email{tsutomu.tyanagida@sjtu.edu.cn}
\affiliation{Tsung-Dao Lee Institute, Shanghai Jiao Tong University, Shanghai 200240, P.\ R.\ China}
\affiliation{Kavli IPMU (WPI), UTIAS, The University of Tokyo, 5-1-5 Kashiwanoha, Kashiwa, Chiba 277-8583, Japan}
\date{\today}

\begin{abstract}
We investigate a quintessence axion model for dynamical dark energy, motivated in part by recent results from the Baryon Acoustic Oscillation (BAO) measurements of the Dark Energy Spectroscopic Instrument (DESI) combined with the cosmic microwave background anisotropies and the latest Type Ia supernovae (SNe Ia) data. By carefully treating the initial conditions and parameter sampling, we identify a preferred parameter space featuring a sub-Planckian axion decay constant and a relatively large axion mass, which naturally avoids the quality problem and remains consistent with the perturbative string conjecture. Our parameter scan also uncovers a trans-Planckian regime of theoretical interest, which is only mildly disfavored even by the strongest constraint. Finally, we discuss the possible connection between this model and the recently reported non-zero rotation of the CMB linear polarization angle, emphasizing the broader cosmological implications and the promising prospects for testing this scenario. We show that an $\mathcal{O}(1)$ electromagnetic anomaly coefficient is preferred by the strongest constraint, which is in full agreement with the minimal quintessence axion model.
\end{abstract}
\maketitle

\section{Introduction}

A growing body of observational evidence suggests that the standard cosmological model, with dark energy (DE) described by a cosmological constant $\Lambda$, may be incomplete. The persistent tension in the measurements of the Hubble constant $H_0$, stemming from the discrepancy between nearby observations~\cite{Riess:2021jrx, Riess:2024vfa} and the rate inferred from early Universe probes within $\Lambda$CDM~\cite{Planck:2018nkj, Planck:2018vyg, ACT:2020gnv, ACT:2025fju, ACT:2025tim, SPT-3G:2022hvq} remains one of the most pressing challenges in cosmology, see Ref.~\cite{CosmoVerse:2025txj} for a recent summary. Different methods of determining the parameter $S_8$ also reveal a potential tension between early- and late-time cosmological observations, with some large-scale structure analyses showing up to a $4.5\sigma$ discrepancy from \textit{Planck} cosmic microwave background (CMB) results~\cite{Ivanov:2024xgb, Chen:2024vuf}, despite recent cosmic shear and other probes agreeing well with $\Lambda$CDM predictions~\cite{Wright:2025xka, Stolzner:2025htz, ACT:2025fju}.

These discrepancies suggest that $\Lambda$CDM may need revision, motivating alternative models, though none have yet resolved all existing tensions. See Refs.~\cite{DiValentino:2020zio, DiValentino:2021izs, Perivolaropoulos:2021jda, Abdalla:2022yfr} for detailed reviews. In particular, cosmological tensions have renewed interest in models where dark energy is not a cosmological constant but instead a dynamical field evolving over cosmic time. One such approach involves parametrizations like the Chevallier-Polarski-Linder (CPL) form~\cite{Chevallier:2000qy, Linder:2002et}, as well as phenomenological frameworks such as quintessence~\cite{Ratra:1987rm, Zlatev:1998tr}. Recent measurements of the baryon acoustic oscillations (BAO) of the Dark Energy Spectroscopic Instrument (DESI)~\cite{DESI:2013agm, DESI:2016fyo, DESI:2016igz} combined with the latest SNe Ia data~\cite{Scolnic:2021amr,Rubin:2023ovl,DES:2024jxu} and Planck CMB power spectra~\cite{Planck:2018vyg} further support this possibility, with the second data release (DR2) hinting at a dynamical nature of DE~\cite{DESI:2025zpo, DESI:2025zgx, DESI:2025fii, Yang:2025kgc} or modified gravity~\cite{Yang:2024kdo,Yang:2025mws}. The higher precision from DESI starts to tighten constraints and challenge the standard $\Lambda$CDM model in some regions of parameter space. Some of these results are in tension with what has been obtained with the Sloan Digital Sky Survey (SDSS)~\cite{SDSS:2002jsq, 2011AJ....142...72E}.

A particularly compelling realization of quintessence involves a pseudoscalar axion-like field that slowly rolls from an initial value~\cite{Choi:2021aze, Visinelli:2018utg, Reig:2021ipa}, contributing to the Universe's total energy density. To match the observed DE today, this field is expected be extremely light, with a mass of order $m_A \simeq H_0 \sim 10^{-33}\,{\rm eV}$. A well-motivated candidate for such a quintessence field is a Nambu–Goldstone boson (NGB) resulting from the spontaneous breaking of a global $U(1)_X$ symmetry~\cite{Fukugita:1994hq, Frieman:1995pm, Choi:1999xn, Ng:2000di, Nomura:2000yk, Kawasaki:2001bq, Caldwell:2005tm, Dutta:2006cf}. In the regime $m_A \simeq H_0$ the dynamics of the field become indistinguishable from those of a cosmological constant. Conversely, if $m_A \gg H_0$ and the field is initially not at nearly potential top, the field tends to behave like pressureless matter~\cite{Riotto:2000kh, Amendola:2005ad, dePutter:2008wt, Hlozek:2014lca}. Ultra-light axion DE models have been tested in the literature using CMB data from \textit{Planck}, ACT and SPT~\cite{Hlozek:2014lca}, as well as the galaxy power spectrum~\cite{Lague:2021frh} and bispectrum ~\cite{Rogers:2023ezo} measurements from the Baryon Oscillation Spectroscopic Survey (BOSS). Recently, the joint analysis of the latest BAO and SNe Ia data combined with \textit{Planck} shows that quintessence models provide a slightly better fit than a cosmological constant~\cite{Berghaus:2024kra, Tada:2024znt, Notari:2024rti, Luu:2025fgw, Shlivko:2024llw, Urena-Lopez:2025rad, Silva:2025hxw,Gialamas:2024lyw}, although they may worsen the Hubble tension~\cite{Banerjee:2020xcn,Lee:2022cyh}, and could be in tension with DESI full-shape measurements~\cite{Colgain:2025nzf}. Nevertheless, considering a quintessence field remains essential for addressing fundamental questions related to the preferred mass scale of the light field, the scale of the decay constant, and its possible connection to the gravitational scale. We emphasize the presence of a class of high-quality quintessence axion models with large decay constants $F_A> 10^{16}$\,GeV~\cite{Girmohanta:2023ghm}, as shown in Appendix~\ref{sec:model}.

Besides the aspects above, the quintessence axion connects with the observational hints of parity-violating physics. Indeed, measurements of CMB anisotropies and related observables may offer insights into deviations from $\Lambda$CDM. Notably, recent analyses of the CMB polarization suggest a non-zero cosmic birefringence angle, $\beta = \mathcal{O}(0.1)\,$deg~\cite{Minami:2020odp, Diego-Palazuelos:2022dsq, Eskilt:2022wav, Eskilt:2022cff,ACT:2025fju}. If confirmed, this rotation could signal the existence of physics beyond SM, potentially arising from parity violation induced by a light scalar field coupled to electromagnetism. The observed birefringence can be induced by a non-zero field excursion between today and the time of last scattering, in models where the axion couples to the SM photon~\cite{Fujita:2020ecn, Takahashi:2020tqv, Fung:2021wbz, Nakagawa:2021nme, Jain:2021shf, Choi:2021aze, Murai:2022zur, Eskilt:2023nxm, Nakagawa:2025ejs,Lee:2025yvn}. For realistic values of the anomaly coefficient, the rotation angle $\beta$ and the field excursion in units of the axion decay constant $F_A$ are of comparable magnitude.

In this work, we explore a quintessence axion as a candidate for dark energy. We assume that the quintessence axion fully accounts for the dark energy observed, setting the cosmological constant to zero. Building upon previous studies, we show that a suitable treatment of the initial conditions reveals a preference for an axion decay constant well below the Planck scale. This scenario is free from the quality problem and remains consistent with the perturbative string conjecture. Although only mildly favored by current data than the regime with a trans-Planckian decay constant, it corresponds to a relatively large axion mass compared to the Hubble constant and yields a consistent cosmological evolution. At the same time, our parameter scan also uncovers a region with a trans-Planckian decay constant, which may offer new insights from a theoretical standpoint. This result remains robust when DESI BAO measurements are considered in combination with other CMB and supernovae datasets. When assessing the model's validity against cosmic birefringence results, we find a best-fit value for the anomalous coupling between the quintessence axion and the photon can be $\mathcal{O}(1)$, resulting from a large field excursion with a sub-Planckian decay constant but yet producing a consistent cosmic expansion history.

%that is significantly lower than the value we previously obtained in Ref.~\cite{Choi:2021aze} using \textit{Planck}+SDSS data. This is consistent with the smaller value of $F_A$ required to accommodate the DESI data within the theoretical framework.

The paper is organized as follows. The datasets and the prior considered are presented in Section~\ref{sec:methods}. The results are organized in Section~\ref{sec:results} and a discussion is drawn in Section~\ref{sec:discussion}. Finally, we present our conclusions in Section~\ref{sec:conclusions}. We review a high-quality quintessence axion model in Appendix~\ref{sec:model}. We work in natural units with $\hbar = c = 1$.

\section{Methods}
\label{sec:methods}

\subsection{Datasets}
In this analysis, we make use of three categories of cosmological data, similar to those adopted in Ref.~\cite{DESI:2025zgx}: baryon acoustic oscillations (BAO), compressed constraints from cosmic microwave background (CMB), and SNe Ia. The BAO constraints are obtained from the second data release (DR2) of the Dark Energy Spectroscopic Instrument (DESI)~\cite{DESI:2025zgx}, providing precise BAO measurements across a wide redshift range. For the CMB, instead of the full temperature and polarization power spectra, we adopt a set of compressed information, which makes use of the CMB information by incorporating the temperature, polarization, and temperature–polarization cross-correlation power spectra from \textit{Planck}~\cite{Planck:2018vyg}. %and the combined \textit{Planck} and the Atacama Cosmology Telescope Data Release 6 (ACT - DR6) CMB lensing data~\cite{ACT:2023kun}. 
This includes the angular scale of the sound horizon at recombination $\theta_*$, the reduced baryon density $\Omega_b h^2$, and the reduced total matter density $\Omega_{\rm m} h^2$ with the following mean ~\cite{Lemos:2023xhs}
\begin{equation}\label{eq:CMB-compressed-means}
    \bm{\mu}(\theta_*,~\Omega_b h^2,~\Omega_{\rm m} h^2)=\begin{pmatrix}0.01041\\0.02223\\0.14208\end{pmatrix}
\end{equation}
and covariance
\begin{equation}\label{eq:CMB-compressed-cov}
    \bm{C}=10^{-9}\begin{pmatrix}0.006621 & 0.12444 & -1.1929\\0.12444 & 21.344 & -94.001\\-1.1929 & -94.001 & 1488.4\end{pmatrix}\,.
\end{equation}
These are incorporated into our joint analysis through a Gaussian likelihood.
These three parameters capture the most relevant information for our purposes (e.g., constraining a late-time DE model) while significantly reducing computational complexity. We incorporate two independent compilations of SNe Ia data: the PantheonPlus sample~\cite{Scolnic:2021amr} and the Dark Energy Survey (DES) Y5 dataset~\cite{DES:2024jxu}, which are used separately. To distinguish between the different SNe Ia data, we label the full combinations as DESI + CMB + PantheonPlus and DESI + CMB + DESY5. 

The compressed CMB likelihood approach follows earlier work~\cite{DESI:2025zgx}, where their left panel of Fig. 14 demonstrated that the small set of parameters listed above can successfully reproduce the dominant \textit{Planck} constraints relevant to late-time cosmology~\cite{Prince:2019hse}. While the mean and covariance given in Eqs.~\eqref{eq:CMB-compressed-means} and~\eqref{eq:CMB-compressed-cov} are derived assuming the standard $\Lambda$CDM model, it has been shown in Ref.~\cite{Lemos:2023xhs} that these statistics remain highly stable across the late-time cosmological models explored in Ref.~\cite{Planck:2018vyg}. This stability is expected: $\theta_*$ is extracted directly from the location of the acoustic peaks in the CMB spectrum, while $\Omega_{\rm b}h^2$ and $\Omega_{\rm m}h^2$ are constrained by the scale dependence and peak ratios of the CMB power spectra~\cite{Hu:2000ti,Planck:2018vyg}. These observable features are insensitive to modifications of late-time cosmology, as long as the physical meanings of $\Omega_{\rm b}h^2$ and $\Omega_{\rm m}h^2$ remain consistent across cosmic time. Thus, although approximate, this method offers a computationally efficient alternative that captures the most relevant information of full CMB code analyses in studies of extended late-time cosmological models. 

A caveat of the compressed CMB approach is that certain information—such as gravitational lensing and other secondary anisotropies—is not captured. While including these effects could lead to slightly tighter constraints, we do not expect them to significantly alter our results. Similarly, the late-time BAO measurements from DESI are also provided in a compressed form, and we have not included the full-shape matter power spectrum in our analysis. Also, similar to Ref.~\cite{Urena-Lopez:2025rad}, we ignore the negligible density perturbations of the quintessence axion field; see also references therein for further discussion.

For the likelihoods corresponding to the three observations above, we largely follow the treatments in Ref.~\cite{Lin:2021sfs}, with one modification: instead of treating the sound horizon uncalibrated, we compute the normalized sound horizon scales at recombination and at the drag epoch assuming the standard cosmological model applies prior to recombination. Specifically, we calculate the normalized sound horizon as
\begin{equation}\label{eq:normalized-sound-horizons}
    r_{\rm s}(z_s)H_0=\int^{z_s}_{\infty}\frac{c_{\rm s}(z)}{\sqrt{\Omega_{\rm m}(1+z)^3+\Omega_{\rm r}(1+z)^4}}dz\,,
\end{equation}
where $c_{\rm s}$ is the sound speed given by
\begin{equation}\label{eq:sound-speed}
    c_{\rm s}(z;\,\Omega_{\rm b}h^2)=\frac{1}{\sqrt{3\big(1+\frac{1}{1+z}\frac{3\,\Omega_{\rm b}h^2}{4\,\Omega_{\gamma}h^2}\big)}}\,.
\end{equation}
Here, we neglect the contribution from dark energy at early times, for redshifts $z \gg 1$. The redshifts $z_s$ correspond to  $z_*$ (recombination) and $z_{\rm d}$ (drag epoch), for which we adopt the fitting formulas from Ref.~\cite{Hu:1995en}. Note that the likelihoods in Ref.~\cite{Lin:2021sfs} are based on the $\Lambda$CDM model post-recombination, and are extended to accommodate non-standard cosmologies~\cite{Wang:2025mqz}. 
We note that the absolute magnitude of SNe Ia is degenerate with the Hubble constant when the former is not independently calibrated. Therefore, following Ref.~\cite{Lin:2021sfs}, we combine these quantities into a single free parameter, defined as:
\begin{equation}\label{eq:C-H-M}
\mathcal{M} \equiv M_0 - 5\log_{10}(10\,{\rm pc} \times H_0)\,.
\end{equation}
This parameter is constrained along with other cosmological parameters and marginalized over at the end of the analysis.

\subsection{Priors and parameter sampling strategy}

We first discuss the cosmological evolution of the field $A$. Acting as a quintessence dark energy candidate, the field slowly rolls from near the top of its potential, starting from an initial configuration $A_i$. In the case of a quintessence axion, the potential energy is typically modeled by a periodic cosine function of the form $V(A) = (V_0/2) \left[1 - \cos\left(A/F_A\right) \right]$, where $F_A$ is the effective axion decay constant and $V_0$ sets the energy scale associated with non-perturbative symmetry-breaking effects. See Eq.~\eqref{eq:selfpotential} for an explicit model build. The model features two free parameters: the quintessence axion mass, given by $m_A^2 = V_0/(2F_A^2)$ in Eq.~\eqref{eq:selfpotential}, and the initial displacement from the potential maximum,
\begin{equation}
    \label{def:delta}
    \delta_i \equiv \pi - A_i / F_A\,.
\end{equation}
We define the dimensionless decay constant ratio as
\begin{equation}
    \label{eq:kappa}
    \kappa_f \equiv F_A / M_P\,.
\end{equation}
This parameter can be determined by specifying $m_A$ and $\delta_i$. For details on the numerical solution of the axion dynamics, see Appendix~\ref{sec:EOM-numerical}.

In addition to the other standard cosmological parameters, the quintessence axion model depends on the two parameters $m_A/H_0$ and the initial displacement $\delta_i$ in Eq.~\eqref{def:delta}. To constitute a viable DE model, these two parameters must satisfy a tight correlation. Specifically, a larger value of $m_A / H_0$, corresponding to a smaller parameter $\kappa_f$, necessitates a smaller $\delta_i$. Based on a WKB approximation, the required $\delta_i$ decreases rapidly as $\kappa_f$ becomes smaller, following the relation $\delta_i \sim \exp(-1/\kappa_f)/\kappa_f$~\cite{Ibe:2018ffn,Lin:2022khg}. Motivated by this, we further perform a numerical analysis and find that the relation is better approximated by
\begin{equation}
    \label{eq:deltai-kf-estimate}
    \delta_i \sim \frac{1}{\kappa_f^{1.5}}\,\exp\left(-\frac{1}{\kappa_f}\right)\,.
\end{equation}
For instance, when $\kappa_f = 0.1$, this expression yields $\delta_i \approx 10^{-3}$. 

Such a stringent requirement on $\delta_i$ introduces an artificial lower bound on $\kappa_f$ or, equivalently, an upper bound on $m_A / H_0$, under certain parameter sampling strategies during parameter inference. This effect arises particularly: (1) when the small-$\delta_i$ region occupies only a small fraction of the prior volume, such as when using a uniform prior over a linear-$\delta_i$ scale, which compresses the viable parameter space into a narrow region at low $\kappa_f$; or (2) when a hard lower bound on $\delta_i$ is explicitly imposed. In both cases, the resultant upper bound on $m_A/H_0$ is not observationally driven but instead emerges as a sampling artifact.

To avoid this sampling bias, motivated by Eq.~\eqref{eq:deltai-kf-estimate} and the approximation $\kappa_f \sim H_0/m_A$ in the small-$\delta_i$ limit of Eq.~\eqref{eq:eqn-for-kf}, we redefine the parameterization using a transformation from $\delta_i$ to the quantity $\epsilon$ as:
\begin{equation}
    \label{eq:epsilon-definition}
    \delta_i = \epsilon\left(\frac{m_A}{H_0}+1\right)^{1.5}\exp\left(-\frac{m_A}{H_0}\right)\,.
\end{equation}
The ‘+1’ ensures that $\epsilon$ remains of order unity when $m_A/H_0$ is small.
We then perform parameter sampling uniformly in $\epsilon$, which ensures proper coverage of the small-$\delta_i$ region. Notably, the $\epsilon = 0$ case recovers the cosmological constant scenario, as the field $A$ with this condition is non-dynamical and sits at the top of the cosine potential.

This transformation effectively avoids sampling outside the DE regime (e.g., scenarios where the quintessence axion field $A$ oscillates at late times). To further enforce the interpretation of the model as dark energy, we impose an additional requirement: the derivative of the field $A'$ with respect to the logarithm of the scale factor $a \equiv 1/(1+z)$ must be greater at present ($a = 1$) than any prior times. This ensures that the field $A$ has not yet reached the potential minimum. In addition, we restrict our analysis to the region where $\kappa_f < 10$, or equivalently $F_A < 10\,M_P$, acknowledging that part of this region may violate the perturbative string conjecture. Larger values of $\kappa_f$ are not necessary to include, as for $\kappa_f \gg 1$, the field begins near the minimum of its potential. In this limit, the potential effectively becomes quadratic, $V \to \frac{1}{2} m_A^2 A^2$, and the dynamics become independent of $F_A$. Consequently, the observational constraints asymptotically converge at large  $F_A$. The parameter ranges used in our analysis are summarized in Table~\ref{tab:parameter-prior}.
\begin{table}[tbp]
    \centering
    \begin{tabular}{l @{\hspace{1.5cm}} l}
    \hline
    Parameter & Uniform Prior \\
    \hline
    $\Omega_{\rm m}$ & $[0.01,\ 0.99]$ \\
    $\log_{10}(m_A / H_0)$ & $[-0.5, 1.3]$ \\
    $\epsilon$ & $[0,\ 25]$ \\
    $h$ & $[0.55,\ 0.88]$ \\
    $\Omega_{\rm b} h^2$ & $[0.01,\ 0.04]$ \\
    $\mathcal{M}$ & $[17,\ 30]$ \\
    \hline
    \end{tabular}
    \caption{Uniform prior ranges for the cosmological parameters and SNe Ia parameter $\mathcal{M}$ used in the analysis. In addition to these priors, we impose the condition that the field derivative $A'$ at $a = 1$ must be greater than at any earlier time. We additionally restrict $F_A/M_P<10$, beyond which the constraints exhibit asymptotic independence from $F_A$.}
    \label{tab:parameter-prior}
\end{table}
In addition to the fiducial analysis using the ``$\epsilon$-sampling'' method described above, we also perform an analysis based on uniform sampling in $\delta_i\in[0,\pi]$, referred to as ``$\delta_i$-sampling''. This alternative approach yields results similar to those reported in Ref.~\cite{Urena-Lopez:2025rad}.

\section{Results}
\label{sec:results}

We solve numerically the equation of motion for the axion dynamics, following the details described in Appendix~\ref{sec:EOM-numerical}. We perform parameter inference via MCMC simulations using the python package \texttt{emcee}~\cite{emcee} with the two sampling methods introduced in Section~\ref{sec:methods}, allowing us to explore the posterior distributions and assess the constraints on the model parameters.

\begin{figure}[tbp]
\centering
\includegraphics[width=0.48\textwidth]{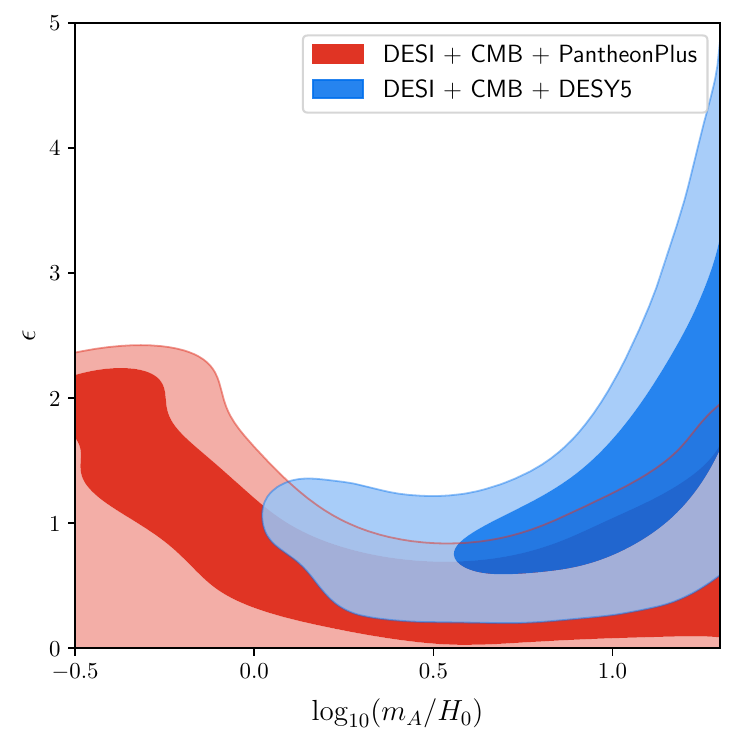}
\caption{Constraints in the  $m_A / H_0$  versus  $\epsilon$ plane from our fiducial analysis using the $\epsilon$-sampling method. This sampling strategy ensures adequate exploration of the small-$m_A$  region, as illustrated in the plot.}
\label{fig:epsilon-sampling}
\end{figure}

Fig.~\ref{fig:epsilon-sampling} presents the constraints in the $m_A / H_0$ versus $\epsilon$ parameter space obtained under our fiducial analysis, i.e., using the $\epsilon$-sampling method. Across most range of $m_A$ considered, the $\epsilon = 0$ limit, which corresponds to dynamics effectively indistinguishable from a cosmological constant, is excluded at $1\sigma-2\sigma$ levels, especially for the results with DESY5. This finding is consistent with the fact that, within the $w_0$–$w_a$ parametrization, the $\Lambda$CDM model is also disfavored at a similar confidence level by the combined DESI+\textit{Planck}+SNe Ia datasets~\cite{DESI:2025zgx}. The data also impose a lower bound of $m_A / H_0 \gtrsim 1$. Values below this are disfavored, as they correspond to field variation timescales that are too long, again effectively mimicking the behavior of a cosmological constant.

\begin{figure}[t]
    \centering
    \includegraphics[width=0.48\textwidth]{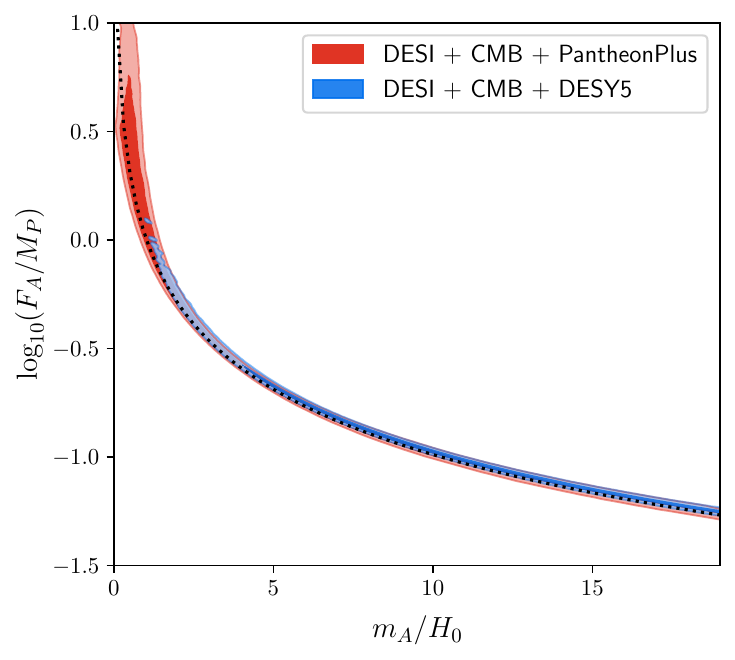}
    \includegraphics[width=0.48\textwidth]{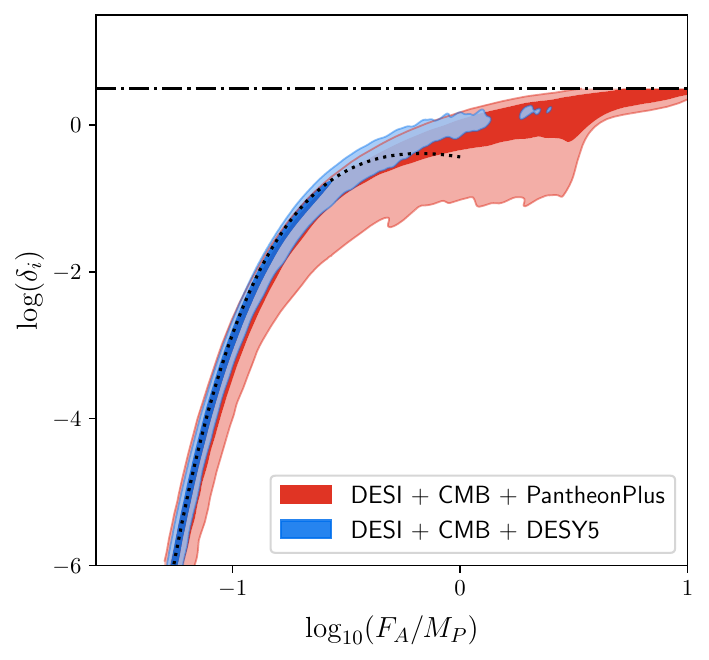}
    \caption{Constraints on the $m_A/H_0$ versus $F_A/M_P$ plane (upper panel) and the $\delta_i$ versus $F_A/M_P$ plane (lower panel) from our fiducial analysis. Upper: The dotted curve shows the relation in Eq.~\eqref{eq:fit}. Lower: The dotted curve shows the relation in Eq.~\eqref{eq:deltai-kf-estimate}, when the equal sign is restored. The dotted-dashed line shows $\delta_i=\pi$, which corresponds to the limit where the quintessence axion potential reduces to a quadratic potential.}
    \label{fig:ma_H0-kf}
\end{figure}

An important parameter in the analysis is $F_A$. The upper panel of Fig.~\ref{fig:ma_H0-kf} shows constraints in both the $m_A/H_0$ versus $F_A/M_P$ and the $\delta_i$ versus $F_A/M_P$ planes, based on the $\epsilon$-sampling method. The dotted curve illustrates the relation between $F_A/M_P$ and $m_A/H_0$ in the small-$\delta_i$ limit of Eq.~\eqref{eq:eqn-for-kf}, given by
\begin{equation}
\label{eq:fit}
\frac{F_A}{M_P} = \frac{3}{2}(1 - \Omega_{\rm m})\frac{H_0}{m_A},
\end{equation}
taking $\Omega_{\rm m} = 0.3$. We can see that as long as $F_A/M_P$ and $m_A/H_0$ approximately follow this relation, the quintessence axion DE provides a comparably good fit to the data. However, the analysis favors values $m_A/H_0\gtrsim 1$ especially for the constraint form DESI+CMB+DESY5, corresponding to a lower energy scale $F_A/M_P \sim 1$ or below, the extent of which depends on the specific dataset combination. Nonetheless, this mild preference for smaller $F_A$ is not statistically strong; see Section~\ref{sec:discussion} for further discussion.

The lower panel of Fig.~\ref{fig:ma_H0-kf} displays the constraints in the $\delta_i$ versus $F_A/M_P$ plane. As expected, the required value of $\delta_i$ decreases sharply as $F_A/M_P$ decreases. However, as long as the parameters approximately satisfy Eq.~\eqref{eq:deltai-kf-estimate}, the quintessence axion DE remains consistent with the data.

It is somewhat surprising that the large-mass regime (and consequently small $F_A$) can provide a good fit to the data. Intuitively, a large axion mass would drive the derived equation-of-state parameters, $w_0$ and $w_a$, to values that significantly deviate from those of the $\Lambda$CDM case. However, $w_0$ and $w_a$ are not directly observable quantities. The key observable is the comoving distance as a function of redshift. For example, under the best-fit parameters, both a low-mass scenario (e.g., $m_A/H_0 = 3$) and a high-mass scenario (e.g., $m_A/H_0 = 15$) yield nearly identical comoving distance–redshift relations, despite having markedly different equation-of-state behaviors. This highlights that the $w_0$–$w_a$ parametrization does not accurately capture the physical behavior of quintessence axion dark energy, particularly in the high-mass regime. Recent studies have explored dark energy models with a hilltop quadratic potential, which corresponds to the leading order of a cosine-type potential~\cite{Bhattacharya:2024kxp,Wolf:2023uno,Wolf:2024eph}. It is important to emphasize that this approximation does not accurately capture the physical behavior in both the large decay constant regime ($F_A/M_P \gtrsim 1$), where the initial field displacement is large, and the small decay constant regime ($F_A/M_P \lesssim 0.5$), where the field excursion must be of order $\mathcal{O}(1)$ (see also Section \ref{sec:CB}). Nevertheless, a mild preference for the larger-mass regime was similarly observed using this approximation with older data~\cite{Wolf:2024eph}.

\begin{figure} 
    \centering
    \includegraphics[width=0.48\textwidth]{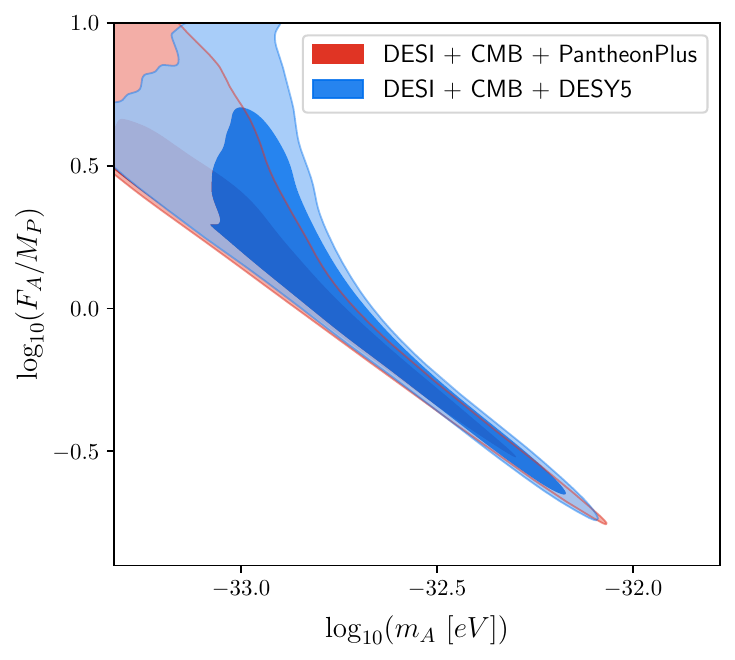}
    \caption{Results using a uniform $\delta_i$ sampling with $0.01<\delta_i<\pi$. The lower limit of $m_A$ shown here is only set by the parameter sampling method.}
    \label{fig:delta-sampling}
\end{figure}

We also investigate the impact of adopting linear-$\delta_i$ sampling. To illustrate this, Fig.~\ref{fig:delta-sampling} shows the results obtained using the $\delta_i$-sampling method with a uniform prior over the range $[0.01,\pi]$. As discussed in Section~\ref{sec:methods}, this sampling scheme introduces a bias toward larger values of the axion decay constant $F_A$, effectively excluding the region $F_A \ll M_P$, which is of particular interest from a model-building perspective. Despite differences in the adopted priors and the treatment of CMB data, the results shown in Fig.~\ref{fig:delta-sampling} are broadly consistent with those reported in Ref.~\cite{Urena-Lopez:2025rad}. Since this exclusion of the small-$F_A$ regime arises from the choice of sampling method rather than observational constraints, our fiducial results are based on the $\epsilon$-sampling method.

\section{Discussion}
\label{sec:discussion}

\subsection{Range of the axion decay constant}

Our analysis reveals a mild preference for larger axion masses, corresponding to smaller values of the decay constant $F_A$. Specifically, for the combined datasets of DESI+CMB+DESY5, we observe that the best-fit $\chi^2$ decreases gradually by about 3 as $m_A/H_0$ increases from 0.5 to 20. For the combined datasets of DESI+CMB+PantheonPlus, the best-fit $\chi^2$ also slightly drops by 1 for the same mass range.  This trend suggests that axion models with higher mass and correspondingly lower $F_A$ provide a slightly better fit to the data. In the regime of large decay constants, particularly for $F_A \gtrsim M_P$ and $m_A/H_0 \lesssim1$, the $\chi^2$ becomes nearly independent of $F_A$. This behavior indicates that current data are largely insensitive to the precise value of $F_A$, and data do not place strong constraints on its value. 

On one hand, a sub-Planckian $F_A$ is theoretically advantageous. The results presented in this work support models with $F_A \sim 10^{17}$\,GeV, such as the one proposed in Ref.~\cite{Girmohanta:2023ghm}, which simultaneously address the quality problem, mitigate fine-tuning, and remain consistent with observational data; further details can be found in Appendix \ref{sec:model}. 

On the other hand, while small values of $F_A$ are slightly favored by the data, scenarios with trans-Planckian decay constants remain consistent with the data, although with marginally higher $\chi^2$. This is remarkable given the conventional expectation that the spontaneous breaking of global symmetries, such as the one associated with the QCD axion, should occur below the Planck scale, an assumption influenced by string-theoretic conjectures~\cite{Arkani-Hamed:2006emk} (see also Ref.~\cite{Li:2025cxn}). The fact that our results do not exclude the possibility of $F_A > M_P$ challenges this conventional viewpoint and opens the door to alternative theoretical scenarios. It may motivate a re-examination of symmetry-breaking mechanisms in ultraviolet completions of axion models and encourage further investigation into how effective global symmetries could persist at trans-Planckian scales.

\subsection{Cosmic birefringence}\label{sec:CB}

\begin{figure}[tbp]
    \centering
    \includegraphics[width=0.48\textwidth]{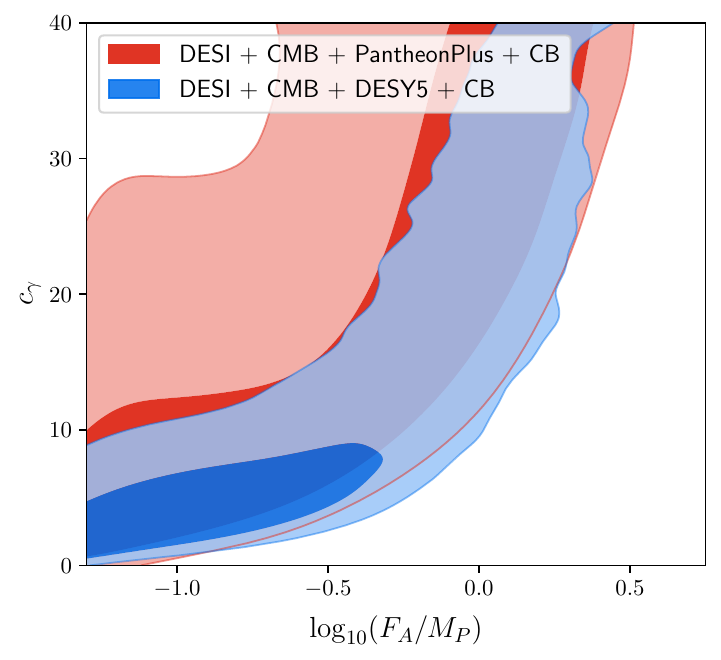}
    \caption{Constraints on the $c_\gamma$ versus $F_A/M_P$ plane when including CB in the analysis. Notably, for DESI + CMB + DESY5 + CB, an electromagnetic anomaly coefficient $c_\gamma$ of $\mathcal{O}(1)$ is preferred, which is consistent with the prediction of the minimal model.}
\label{fig:cy}
\end{figure}

Quintessence axion predicts a notable phenomenon known as cosmic birefringence (CB)–a rotation of the polarization angle of photons traveling over cosmological distances. CB is of particular interest as it provides a potential signal of parity violation on cosmological scales. The theoretical relation between the rotation angle $\beta$ and the field excursion, arising from the quintessence axion–photon coupling described in Eq.~\eqref{eq:CS}, is given by~\cite{Carroll:1989vb, Carroll:1991zs, Harari:1992ea}:
\begin{equation}
\label{eq:beta}
 \beta=0.42{\rm\, deg}\,\left(\frac{c_\gamma}{2\pi}\right)\,\left(\frac{A_0-A_{\rm LSS}}{F_A}\right)\,,
\end{equation}
where $c_\gamma = \mathcal{O}(1 \textrm{--} 10)$ is the electromagnetic anomaly coefficient, see Eq.~\eqref{eq:CS}, $A_{\rm LSS}$ is the field value at the surface of last scattering, and $A_0$ is its present-day value. After solving the dynamics of the axion field $A$, we compute $\beta$ consistently for the same model parameters.

Intriguingly, there is now observational evidence suggesting a non-vanishing $\beta$. Based on \textit{Planck} Data Release 4, it has been reported at the 68\% confidence level that~\cite{Minami:2020odp, Diego-Palazuelos:2022dsq, Eskilt:2022wav, Eskilt:2022cff}:
\begin{equation}
\beta_{\rm Planck} = 0.342^{+0.094}_{-0.091}\,{\rm deg}\,.
\end{equation}
A consistent measurement was recently reported by ACT, also at the 68\% confidence level~\cite{ACT:2025fju}:
\begin{equation}
\beta_{\rm ACT} = 0.20\pm0.08\,{\rm deg}\,.
\end{equation}
Assuming statistical independence between the two datasets, we combine the measurements to obtain\footnote{See e.g.\ Section 40.~\textit{Statistics} in The Particle Data Group 2022 review~\cite{ParticleDataGroup:2022pth}}
\begin{equation}
    \beta = 0.261 \pm 0.061\,{\rm deg}\,.
\end{equation}
Although the \textit{Planck} and ACT DR6 datasets cover overlapping regions of the sky, the multipole ranges that dominate their respective birefringence constraints are largely complementary. The \textit{Planck} analysis is primarily sensitive to large angular scales, with most of the signal-to-noise coming from $\ell \lesssim 800$, while the ACT analysis relies on small-scale measurements, with the dominant contribution arising from $\ell \gtrsim 800$. Due to this limited overlap in multipole space, we approximate the two datasets as statistically independent when combining their constraints.\footnote{Recently, the work in Ref.~\cite{Ballardini:2025apf} appeared, providing an independent indication for a non-zero cosmic birefringence in CMB photons.}

We perform a new analysis by incorporating an additional constraint from the axion field excursion required to explain the observed CB, using the $\epsilon$-sampling method. In this analysis, the electromagnetic anomaly coefficient $c_\gamma$ is treated as a new free parameter. The resulting constraint is shown in Fig.~\ref{fig:cy}. We find that the required value of $c_\gamma$ decreases with increasing $m_A/H_0$, and consequently with decreasing $F_A/M_P$.

Remarkably, for $m_A/H_0 \gtrsim 5 $, or $F_A/M_P \lesssim 0.2$, the anomaly is required to be $c_\gamma = \mathcal{O}(1)$. Previous studies, including our own work~\cite{Choi:2021aze}, reported relatively large values of the electromagnetic anomaly coefficient, $c_\gamma \sim \mathcal{O}(20)$, across the full range of $F_A$. In contrast, the results presented in Fig.~\ref{fig:cy} favor lower values of $c_\gamma$, especially for the data combination DESI + CMB + DESY5 + CB, corresponding to a lower preferred range for $F_A$. This difference stems \textbf{not} from the improvement of data quality, but rather from the improved treatment of the small-$F_A$ regime in the present analysis. Unlike previous approaches that relied on indirect equation-of-state parameters, which is misleading due to large field excursions with small $F_A$ still yield a distance-redshift relation consistent with observations, our analysis performs a direct comparison between the comoving distance-redshift relation and observational data.
Since $c_\gamma$ originates from a specific combination of gauge group anomalies, smaller values are theoretically more attractive, particularly in the context of ultraviolet-complete models. This provides sharper guidance for model building.

\section{Conclusions}
\label{sec:conclusions}

In this work, we have explored a quintessence axion model for dynamical dark energy assuming a vanishing cosmological constant, inspired in part by recent BAO measurements from DESI DR2, and CMB observations from ACT. By carefully modeling and sampling the initial conditions in the analysis, we have identified a preferred region of parameter space featuring a sub-Planckian axion decay constant. This regime naturally avoids the quality problem and remains consistent with the perturbative string conjecture, while predicting a relatively large axion mass compared to the Hubble scale. Additionally, our parameter scan reveals a trans-Planckian regime that, while less phenomenologically favored, presents a theoretically interesting avenue for further investigation. The overall results are robust under the inclusion of DESI BAO data in combination with CMB and SNe Ia datasets. 

Lastly, we have discussed a possible link between the model and the recently reported non-zero cosmic birefringence, observed as a rotation of the CMB linear polarization angle. A minimal setup with an electromagnetic anomaly coefficient $c_\gamma = \mathcal{O}(1)$ and a sub-Planckian axion decay constant can simultaneously account for the observed cosmic birefringence and the late-time cosmic acceleration, is favored at the $1\sigma$ level for the DESI + CMB + DESY5 dataset combination. However, the result becomes less conclusive when the PantheonPlus SNe Ia dataset is included. While a larger decay constant would require a proportionally larger anomaly coefficient, cosmic birefringence may provide an independent observational probe, potentially helping to break the degeneracy in the distance–redshift relation present in the results. Although this connection remains speculative, it offers a compelling framework for interpreting the broader cosmological role of axion quintessence and presents a promising avenue for testing such models.

\begin{acknowledgments}
We thank Eiichiro Komatsu for crucial insights on cosmic birefringence analysis. We further thank Eleonora Di Valentino and Hanyu Cheng for discussions on cosmological simulations. L.\ V.~acknowledges support by the National Natural Science Foundation of China (NSFC) through the grant No.\ 12350610240 ``Astrophysical Axion Laboratories'', as well as Istituto Nazionale di Fisica Nucleare (INFN) through the ``QGSKY'' Iniziativa Specifica project. W. L. acknowledges that this work is supported by the ``Science \& Technology Champion Project" (202005AB160002), the ``Top Team Project" (202305AT350002) and the ``Innovation Team Project'' (202105AE160021), all funded by the ``Yunnan Revitalization Talent Support Program." W. L. is also supported by the ``Yunnan Key Laboratory of Survey Science" (202449CE340002), the ``Yunnan General Grant'' (202401AT070489) and the National Key R\&D Program of China (2024YFA1611600).
This work is in part supported by MEXT KAKENHI Grants No.~24H02244 (T.\ T.\ Y.). T.\ T.\ Y.~is supported also by the Natural Science Foundation of China (NSFC) under Grant No.~12175134 as well as by World Premier International Research Center Initiative (WPI Initiative), MEXT, Japan. 
\end{acknowledgments}

\appendix

\section{High-Quality Quintessence Axion}
\label{sec:model}

We briefly review the quintessence axion model of high quality proposed in Ref.~\cite{Girmohanta:2023ghm}. The model is based on a five dimensional space-time with $S^1/\mathbb{Z}_2$ orbifold compactification and an extra $U(1)_g$ gauge symmetry. We introduce two scalar fields, $\phi_1$ and $\phi_2$, and localize them on separate branes at two distinct fixed points, $\bm{I}$ and $\bm{II}$, respectively. We assume that the $\phi_1$ and $\phi_2$ fields carry a $U(1)_g$ gauge charge equal to $+1$ and $-1$, respectively. We see, immediately, that in the limit where the distance between the two fixed points is infinite, we have an exact global $U(1)_X$ symmetry in addition to the gauge $U(1)_g$ symmetry, as the fields cannot interact with each others. Any explicit breaking of the global $U(1)_X$ symmetry is then forbidden by the gauge $U(1)_g$ symmetry on each branes.

We assume that the two fields, $\phi_1$ and $\phi_2$, acquire vacuum expectation values (vevs) of the form $\langle\phi_i\rangle = f_i/\sqrt{2}$, hence in the vacuum they can be decomposed in terms of the fields $a_i(x)$ as $\phi_i=(f_i/\sqrt{2})\exp(ia_i(x)/f_i)$. For simplicity, we assume the vevs are equal and set $f_1=f_2=f$. In this setup, both gauge and global $U(1)$ symmetries are spontaneously broken. One Nambu-Goldstone (NG) boson is absorbed into the $U(1)_g$ gauge field, while the other remains as a physical, massless NG boson. This physical mode corresponds to the axion, $A=(a_1+a_2)/\sqrt{2}$.

We now discuss the breaking operators for the global $U(1)_X$ symmetry due to bulk dynamics. Ref.~\cite{Girmohanta:2023ghm} considers the exchange of a heavy massive boson $\Phi$ propagating in the bulk, which induces an effective coupling between the two brane-localized bosons, $\phi_1\phi_2$. This interaction explicitly breaks the global $U(1)_X$ symmetry. However, the strength of this operator is exponentially suppressed by a factor $\exp(-M_\Phi L)$, where $L$ is the separation between the two branes. The brane separation $L$ is related to the fundamental scale $M_*$ in the five dimensional space-time as $M_*L=M^2_P/M_*^2$. The mass of the heavy boson $\Phi$ in the bulk is given by $M_\Phi=kM_*$, with $k = \mathcal{O}(1)$. Therefore, we obtain the $U(1)_X$-breaking operator as 
\begin{equation}
    \mathcal{L}_{\rm eff} =M_*^2\phi_1\phi_2e^{-k(M_P/M_*)^2} ~+~\textrm{h.c.}\,,
\end{equation}
where $M_P$ is the reduced Planck mass. This operator generates a potential for the axion $A$, given by
\begin{equation}
    \label{eq:selfpotential}
    V=\frac{V_0}{2}\left[1-\cos\left(\frac{A}{F_A}\right)\right]\,,
\end{equation}
where the amplitude $V_0$ and the axion decay constant $F_A$ are defined as
\begin{equation}
   V_0= 2f^2M_*^2 e^{-k(M_P/M_*)^2}\,,\quad F_A=f/\sqrt{2}\,.
\end{equation}

We take the vacuum expectation values of the scalar fields to be $|\langle\phi_i\rangle| = M_*$, which implies $F_A=M_*$.\footnote{We adopt $|\langle\phi_i\rangle|=M_*/\sqrt{2}$ in Ref.~\cite{Girmohanta:2023ghm}.} The fundamental scale $M_*$ is determined in Ref.~\cite{Girmohanta:2023ghm} from the requirement of reproducing the observed vacuum energy density. As a result, we predict the effective axion decay constant to be\footnote{The effective decay constant $F_A$ can be increased by keeping the same quality if another pair of fields $\phi_{1}$ and $\phi_2$ is introduced~\cite{Lin:2022khg}.}
\begin{equation}
    \label{eq:FACC}
    F_A = M_* \simeq \sqrt{k}\times 1.47\times 10^{17}{\rm \, GeV}\,.
\end{equation}
The high-quality quintessence axion model is an interesting framework for addressing the anthropic principle by explaining the fine-tuning of the cosmological constant: if the axion decay constant $F_A$ is slightly above the value in Eq.~\eqref{eq:FACC}, the resulting vacuum energy becomes too large, conflicting with the entropic principle, while a slightly smaller $F_A$ demands excessively fine-tuned initial conditions. As such, only a very narrow range of $F_A$ yields the observed cosmological constant.

If five dimensional wormhole solutions exist, they can generate the breaking of the global $U(1)_X$  symmetry. However, these wormholes should involve fields located on both branes to break the global $U(1)_X$ symmetry, hence the volumes of the relevant wormhole solutions are enhanced by the distance $L$ between the two branes. As a result, the wormhole effects are exponentially suppressed as $\exp(-k'M_* L) = \exp(-k'M_P^2/M_*^2)$~\cite{Kallosh:1995hi}, where the constant $k'=\mathcal{O}(1)$ is related to the wormhole action.

As suggested in Ref.~\cite{Girmohanta:2023ghm}, we introduce $n$ pairs of electrons and positrons on each brane. These fermions acquire masses via the vevs of the scalar fields $\phi_1$ and $\phi_2$, respectively. As a result, the axion field $A$ acquires an electromagnetic anomaly from one-loop diagrams involving the charged electrons and positrons. This gives rise to the following Chern-Simons coupling term:
\begin{equation}
    \label{eq:CS}
    \mathcal{L}_{\rm CS} = c_\gamma \frac{A}{F_A}\frac{g_{\rm em}^2}{16\pi^2}F_{\mu\nu}\tilde{F}^{\mu\nu},
\end{equation}
where $c_\gamma =n$, $g_{\rm em}$ is the gauge coupling constant associated with the electromagnetic $U(1)_{\rm em}$ gauge symmetry, and $F_{\mu\nu}$ and $\tilde F_{\mu\nu}$ are the electromagnetic tensor and its dual.

\section{Quintessence axion equation of motion and numerical solutions}
\label{sec:EOM-numerical}
\subsection{Differential equations for background evolution}

We consider a spatially flat universe. For the observables considered in this work, it is sufficient to model the background evolution of the quintessence axion field $A$ and the scale factor of the Universe. Following the notation in Ref.~\cite{Choi:2021aze}, the self-interacting potential leads to the time evolution of the quintessence axion field $A$ as
\begin{equation}
    \label{eq:eomofaxion}    
    \ddot{A}+3H\dot{A}+\frac{\partial V(A)}{\partial A}=0\,,
\end{equation}
where $V(A)$ is given in Eq.~\eqref{eq:selfpotential}, $H$ is the Hubble expansion rate, and the dot denotes differentiation with respect to cosmic time. Here, the back-reaction from the axion-photon interaction term in Eq.~\eqref{eq:CS} has been neglected. To facilitate the analysis, we define the following dimensionless variables:
\begin{equation*}
    \kappa_f \equiv \frac{F_A}{M_P}, \quad
    E \equiv \frac{H}{H_0}, \quad
    \widetilde{\Omega}_A \equiv \frac{V_0}{\rho_{\rm c}^0},
\end{equation*} 
where $\rho_{\rm c}^0=3M_P^2H_0^2$ is the critical density today. The last parameter can also be expressed as
\begin{equation}\label{eq:Oa-kf-ma_H0}
\widetilde{\Omega}_A = \frac{2}{3}\kappa_f^2\frac{m_A^2}{H_0^2}.
\end{equation}
Note that $\widetilde{\Omega}_A$ is \textbf{not} the current energy density fraction of the quintessence axion. In general, $\widetilde{\Omega}_A > 1 - \Omega_{\rm m}$, with $\widetilde{\Omega}_A \to 1 - \Omega_{\rm m}$ in the small-$\delta_i$ limit as $A$ would remain on top of the potential such that $V_0$ equals the DE energy density today.

The energy density of $A$ in terms of the displacement $\delta \equiv \pi-A/F_A$ is given by:
\begin{align}\label{eq:energy-density-quintessence-axion}
\rho_A &= V(A) + \frac{1}{2} \dot{A}^2 \nonumber \\
&= m_A^2 F_A^2 (1 + \cos\delta) + \frac{1}{2} H^2 F_A^2 (\delta')^2\,,
\end{align}
where $\delta' \equiv {\rm d}\delta/{\rm d}\ln a$ and the prime denotes differentiation with respect to $\ln a$.

For cosmic times after recombination, we neglect the contribution of radiation to the expansion of the Universe. The Friedmann equation then includes only the matter and quintessence axion contributions:
\begin{equation}\label{eq:1st-Friedmann}
H^2 = \frac{1}{3M_P^2}(\rho_{\rm m} + \rho_A)\,.
\end{equation}
Using Eq.~\eqref{eq:energy-density-quintessence-axion}, we express the Friedmann equation in a dimensionless form:
\begin{equation}\label{eq:E2-total}
    E^2=\frac{\Omega_{\rm m}/a^3+\widetilde{\Omega}_A\frac{1+\cos\delta}{2}}{1-\kappa_f^2(\delta')^2/6}\,.
\end{equation}

With these above, the quintessence axion equation of motion, Eq.~\eqref{eq:eomofaxion}, becomes
\begin{equation}\label{eq:eomofaxion-dimensionless}
    \delta''=\frac{1}{E^2}\left(\frac{m_A^2}{H_0^2}\sin\delta-\frac{3}{2}\Big[\frac{\Omega_{\rm m}}{a^3}+\widetilde{\Omega}_A(1+\cos\delta)\Big]\delta'\right)\,.
\end{equation}
Once the above second-order differential equation for $\delta(a)$ is solved, we substitute the solutions for $\delta(a)$ and $\delta'(a)$ into Eq.~\eqref{eq:E2-total} and evaluate the normalized comoving distance as a function of redshift:
\begin{equation}
    f_M(z)\equiv H_0d_M=\int_0^z\frac{{\rm d}z'}{E(z')}\,,
\end{equation}
which is a key quantity for observables related to standard rulers and standard candles~\cite{Lin:2021sfs}.

\subsection{Initial condition and boundary condition today}

Inspecting Eq.~\eqref{eq:eomofaxion-dimensionless} in the limit $a\to0$ and assuming an asymptotic power-law form $\delta'\to Ca^p$, we find
\begin{equation}\label{eq:initial-delta-prime}
    \lim_{a\to0}\delta'=\frac{2m_A^2\sin\delta_i}{9\Omega_{\rm m}\,H_0^2}a^3\,.
\end{equation}
We integrate Eq.~\eqref{eq:eomofaxion-dimensionless} starting from an arbitrarily small scale factor $a_i$, using a given initial value $\delta_i$ and setting $\delta'(a_i)$ according to Eq.~\eqref{eq:initial-delta-prime}.

The value of $\kappa_f$ is determined by closing the energy budget of the Universe today, i.e., $E(a=1)=1$. From Eq.~\eqref{eq:E2-total}, this gives
\begin{equation}\label{eq:eqn-for-kf}
    \kappa_f=\left(\frac{3(1-\Omega_{\rm m})}{\left(\frac{m_A}{H_0}\right)^2(1+\cos\delta_0)+\frac{1}{2}(\delta'_0)^2}\right)^{1/2}\,,
\end{equation}
where $\delta_0$ and $\delta'_0$ are the present-day values obtained from solving the equation of motion.

For a given pair $(m_A/H_0, \delta_i)$, we determine $\kappa_f$ iteratively. We begin by estimating $\kappa_f$ using Eq.~\eqref{eq:eqn-for-kf} with $\delta_0 = \delta'_0 = 0$, which is a valid approximation in the small-$\delta_i$ limit, which we denote as $\kappa_f^{\rm est}$. Using this estimate, we solve Eq.~\eqref{eq:eomofaxion-dimensionless} to compute $\delta(a)$ and $\delta'(a)$, extract the updated $\delta_0$ and $\delta'_0$, and recalculate $\kappa_f^{\rm est}$ with Eq.~\eqref{eq:eqn-for-kf}. Then, we take the average of the previous and the new $\kappa_f^{\rm est}$'s, put it in Eq.~\eqref{eq:eomofaxion-dimensionless} to solve for $\delta_0 = \delta'_0 = 0$ again. This process is repeated until $\kappa_f^{\rm est}$ converges to within $0.01\%$, which typically requires only a few iterations.

As demonstrated by Eq.~\eqref{eq:deltai-kf-estimate}, in the regime of small $F_A$ or large $m_A$, the initial field displacement $\delta_i$ must be extremely small. This presents a fine-tuning issue for cases with small $F_A$, although it remains relatively mild for the model discussed in Section\ref{sec:model}. Scenarios involving a non-zero initial field velocity can help mitigate this fine-tuning~\cite{Berbig:2024aee}. A detailed exploration of this possibility lies beyond the scope of this work.

\bibliographystyle{apsrev4-1}
\bibliography{ref}

\begin{thebibliography}{98}%
\makeatletter
\providecommand \@ifxundefined [1]{%
 \@ifx{#1\undefined}
}%
\providecommand \@ifnum [1]{%
 \ifnum #1\expandafter \@firstoftwo
 \else \expandafter \@secondoftwo
 \fi
}%
\providecommand \@ifx [1]{%
 \ifx #1\expandafter \@firstoftwo
 \else \expandafter \@secondoftwo
 \fi
}%
\providecommand \natexlab [1]{#1}%
\providecommand \enquote  [1]{``#1''}%
\providecommand \bibnamefont  [1]{#1}%
\providecommand \bibfnamefont [1]{#1}%
\providecommand \citenamefont [1]{#1}%
\providecommand \href@noop [0]{\@secondoftwo}%
\providecommand \href [0]{\begingroup \@sanitize@url \@href}%
\providecommand \@href[1]{\@@startlink{#1}\@@href}%
\providecommand \@@href[1]{\endgroup#1\@@endlink}%
\providecommand \@sanitize@url [0]{\catcode `\\12\catcode `\$12\catcode
  `\&12\catcode `\#12\catcode `\^12\catcode `\_12\catcode `\%12\relax}%
\providecommand \@@startlink[1]{}%
\providecommand \@@endlink[0]{}%
\providecommand \url  [0]{\begingroup\@sanitize@url \@url }%
\providecommand \@url [1]{\endgroup\@href {#1}{\urlprefix }}%
\providecommand \urlprefix  [0]{URL }%
\providecommand \Eprint [0]{\href }%
\providecommand \doibase [0]{http://dx.doi.org/}%
\providecommand \selectlanguage [0]{\@gobble}%
\providecommand \bibinfo  [0]{\@secondoftwo}%
\providecommand \bibfield  [0]{\@secondoftwo}%
\providecommand \translation [1]{[#1]}%
\providecommand \BibitemOpen [0]{}%
\providecommand \bibitemStop [0]{}%
\providecommand \bibitemNoStop [0]{.\EOS\space}%
\providecommand \EOS [0]{\spacefactor3000\relax}%
\providecommand \BibitemShut  [1]{\csname bibitem#1\endcsname}%
\let\auto@bib@innerbib\@empty
%</preamble>
\bibitem [{\citenamefont {Riess}\ \emph {et~al.}(2022)\citenamefont {Riess}
  \emph {et~al.}}]{Riess:2021jrx}%
  \BibitemOpen
  \bibfield  {author} {\bibinfo {author} {\bibfnamefont {A.~G.}\ \bibnamefont
  {Riess}} \emph {et~al.},\ }\href {\doibase 10.3847/2041-8213/ac5c5b}
  {\bibfield  {journal} {\bibinfo  {journal} {Astrophys. J. Lett.}\ }\textbf
  {\bibinfo {volume} {934}},\ \bibinfo {pages} {L7} (\bibinfo {year} {2022})},\
  \Eprint {http://arxiv.org/abs/2112.04510} {arXiv:2112.04510 [astro-ph.CO]}
  \BibitemShut {NoStop}%
\bibitem [{\citenamefont {Riess}\ \emph {et~al.}(2024)\citenamefont {Riess}
  \emph {et~al.}}]{Riess:2024vfa}%
  \BibitemOpen
  \bibfield  {author} {\bibinfo {author} {\bibfnamefont {A.~G.}\ \bibnamefont
  {Riess}} \emph {et~al.},\ }\href {\doibase 10.3847/1538-4357/ad8c21}
  {\bibfield  {journal} {\bibinfo  {journal} {Astrophys. J.}\ }\textbf
  {\bibinfo {volume} {977}},\ \bibinfo {pages} {120} (\bibinfo {year}
  {2024})},\ \Eprint {http://arxiv.org/abs/2408.11770} {arXiv:2408.11770
  [astro-ph.CO]} \BibitemShut {NoStop}%
\bibitem [{\citenamefont {Aghanim}\ \emph
  {et~al.}(2020{\natexlab{a}})\citenamefont {Aghanim} \emph
  {et~al.}}]{Planck:2018nkj}%
  \BibitemOpen
  \bibfield  {author} {\bibinfo {author} {\bibfnamefont {N.}~\bibnamefont
  {Aghanim}} \emph {et~al.} (\bibinfo {collaboration} {Planck}),\ }\href
  {\doibase 10.1051/0004-6361/201833880} {\bibfield  {journal} {\bibinfo
  {journal} {Astron. Astrophys.}\ }\textbf {\bibinfo {volume} {641}},\ \bibinfo
  {pages} {A1} (\bibinfo {year} {2020}{\natexlab{a}})},\ \Eprint
  {http://arxiv.org/abs/1807.06205} {arXiv:1807.06205 [astro-ph.CO]}
  \BibitemShut {NoStop}%
\bibitem [{\citenamefont {Aghanim}\ \emph
  {et~al.}(2020{\natexlab{b}})\citenamefont {Aghanim} \emph
  {et~al.}}]{Planck:2018vyg}%
  \BibitemOpen
  \bibfield  {author} {\bibinfo {author} {\bibfnamefont {N.}~\bibnamefont
  {Aghanim}} \emph {et~al.} (\bibinfo {collaboration} {Planck}),\ }\href
  {\doibase 10.1051/0004-6361/201833910} {\bibfield  {journal} {\bibinfo
  {journal} {Astron. Astrophys.}\ }\textbf {\bibinfo {volume} {641}},\ \bibinfo
  {pages} {A6} (\bibinfo {year} {2020}{\natexlab{b}})},\ \bibinfo {note}
  {[Erratum: Astron.Astrophys. 652, C4 (2021)]},\ \Eprint
  {http://arxiv.org/abs/1807.06209} {arXiv:1807.06209 [astro-ph.CO]}
  \BibitemShut {NoStop}%
\bibitem [{\citenamefont {Aiola}\ \emph {et~al.}(2020)\citenamefont {Aiola}
  \emph {et~al.}}]{ACT:2020gnv}%
  \BibitemOpen
  \bibfield  {author} {\bibinfo {author} {\bibfnamefont {S.}~\bibnamefont
  {Aiola}} \emph {et~al.} (\bibinfo {collaboration} {ACT}),\ }\href {\doibase
  10.1088/1475-7516/2020/12/047} {\bibfield  {journal} {\bibinfo  {journal}
  {JCAP}\ }\textbf {\bibinfo {volume} {12}},\ \bibinfo {pages} {047} (\bibinfo
  {year} {2020})},\ \Eprint {http://arxiv.org/abs/2007.07288} {arXiv:2007.07288
  [astro-ph.CO]} \BibitemShut {NoStop}%
\bibitem [{\citenamefont {Louis}\ \emph {et~al.}(2025)\citenamefont {Louis}
  \emph {et~al.}}]{ACT:2025fju}%
  \BibitemOpen
  \bibfield  {author} {\bibinfo {author} {\bibfnamefont {T.}~\bibnamefont
  {Louis}} \emph {et~al.} (\bibinfo {collaboration} {ACT}),\ }\href@noop {} {\
  (\bibinfo {year} {2025})},\ \Eprint {http://arxiv.org/abs/2503.14452}
  {arXiv:2503.14452 [astro-ph.CO]} \BibitemShut {NoStop}%
\bibitem [{\citenamefont {Calabrese}\ \emph {et~al.}(2025)\citenamefont
  {Calabrese} \emph {et~al.}}]{ACT:2025tim}%
  \BibitemOpen
  \bibfield  {author} {\bibinfo {author} {\bibfnamefont {E.}~\bibnamefont
  {Calabrese}} \emph {et~al.} (\bibinfo {collaboration} {ACT}),\ }\href@noop {}
  {\  (\bibinfo {year} {2025})},\ \Eprint {http://arxiv.org/abs/2503.14454}
  {arXiv:2503.14454 [astro-ph.CO]} \BibitemShut {NoStop}%
\bibitem [{\citenamefont {Balkenhol}\ \emph {et~al.}(2023)\citenamefont
  {Balkenhol} \emph {et~al.}}]{SPT-3G:2022hvq}%
  \BibitemOpen
  \bibfield  {author} {\bibinfo {author} {\bibfnamefont {L.}~\bibnamefont
  {Balkenhol}} \emph {et~al.} (\bibinfo {collaboration} {SPT-3G}),\ }\href
  {\doibase 10.1103/PhysRevD.108.023510} {\bibfield  {journal} {\bibinfo
  {journal} {Phys. Rev. D}\ }\textbf {\bibinfo {volume} {108}},\ \bibinfo
  {pages} {023510} (\bibinfo {year} {2023})},\ \Eprint
  {http://arxiv.org/abs/2212.05642} {arXiv:2212.05642 [astro-ph.CO]}
  \BibitemShut {NoStop}%
\bibitem [{\citenamefont {Di~Valentino}\ \emph {et~al.}(2025)\citenamefont
  {Di~Valentino} \emph {et~al.}}]{CosmoVerse:2025txj}%
  \BibitemOpen
  \bibfield  {author} {\bibinfo {author} {\bibfnamefont {E.}~\bibnamefont
  {Di~Valentino}} \emph {et~al.} (\bibinfo {collaboration} {CosmoVerse}),\
  }\href {\doibase 10.1016/j.dark.2025.101965} {\bibfield  {journal} {\bibinfo
  {journal} {Phys. Dark Univ.}\ }\textbf {\bibinfo {volume} {49}},\ \bibinfo
  {pages} {101965} (\bibinfo {year} {2025})},\ \Eprint
  {http://arxiv.org/abs/2504.01669} {arXiv:2504.01669 [astro-ph.CO]}
  \BibitemShut {NoStop}%
\bibitem [{\citenamefont {Ivanov}\ \emph {et~al.}(2025)\citenamefont {Ivanov},
  \citenamefont {Obuljen}, \citenamefont {Cuesta-Lazaro},\ and\ \citenamefont
  {Toomey}}]{Ivanov:2024xgb}%
  \BibitemOpen
  \bibfield  {author} {\bibinfo {author} {\bibfnamefont {M.~M.}\ \bibnamefont
  {Ivanov}}, \bibinfo {author} {\bibfnamefont {A.}~\bibnamefont {Obuljen}},
  \bibinfo {author} {\bibfnamefont {C.}~\bibnamefont {Cuesta-Lazaro}}, \ and\
  \bibinfo {author} {\bibfnamefont {M.~W.}\ \bibnamefont {Toomey}},\ }\href
  {\doibase 10.1103/PhysRevD.111.063548} {\bibfield  {journal} {\bibinfo
  {journal} {Phys. Rev. D}\ }\textbf {\bibinfo {volume} {111}},\ \bibinfo
  {pages} {063548} (\bibinfo {year} {2025})},\ \Eprint
  {http://arxiv.org/abs/2409.10609} {arXiv:2409.10609 [astro-ph.CO]}
  \BibitemShut {NoStop}%
\bibitem [{\citenamefont {Chen}\ \emph {et~al.}(2024)\citenamefont {Chen},
  \citenamefont {Ivanov}, \citenamefont {Philcox},\ and\ \citenamefont
  {Wenzl}}]{Chen:2024vuf}%
  \BibitemOpen
  \bibfield  {author} {\bibinfo {author} {\bibfnamefont {S.-F.}\ \bibnamefont
  {Chen}}, \bibinfo {author} {\bibfnamefont {M.~M.}\ \bibnamefont {Ivanov}},
  \bibinfo {author} {\bibfnamefont {O.~H.~E.}\ \bibnamefont {Philcox}}, \ and\
  \bibinfo {author} {\bibfnamefont {L.}~\bibnamefont {Wenzl}},\ }\href
  {\doibase 10.1103/PhysRevLett.133.231001} {\bibfield  {journal} {\bibinfo
  {journal} {Phys. Rev. Lett.}\ }\textbf {\bibinfo {volume} {133}},\ \bibinfo
  {pages} {231001} (\bibinfo {year} {2024})},\ \Eprint
  {http://arxiv.org/abs/2406.13388} {arXiv:2406.13388 [astro-ph.CO]}
  \BibitemShut {NoStop}%
\bibitem [{\citenamefont {Wright}\ \emph {et~al.}(2025)\citenamefont {Wright}
  \emph {et~al.}}]{Wright:2025xka}%
  \BibitemOpen
  \bibfield  {author} {\bibinfo {author} {\bibfnamefont {A.~H.}\ \bibnamefont
  {Wright}} \emph {et~al.},\ }\href@noop {} {\  (\bibinfo {year} {2025})},\
  \Eprint {http://arxiv.org/abs/2503.19441} {arXiv:2503.19441 [astro-ph.CO]}
  \BibitemShut {NoStop}%
\bibitem [{\citenamefont {St{\"o}lzner}\ \emph {et~al.}(2025)\citenamefont
  {St{\"o}lzner} \emph {et~al.}}]{Stolzner:2025htz}%
  \BibitemOpen
  \bibfield  {author} {\bibinfo {author} {\bibfnamefont {B.}~\bibnamefont
  {St{\"o}lzner}} \emph {et~al.},\ }\href@noop {} {\  (\bibinfo {year}
  {2025})},\ \Eprint {http://arxiv.org/abs/2503.19442} {arXiv:2503.19442
  [astro-ph.CO]} \BibitemShut {NoStop}%
\bibitem [{\citenamefont {Di~Valentino}\ \emph
  {et~al.}(2021{\natexlab{a}})\citenamefont {Di~Valentino} \emph
  {et~al.}}]{DiValentino:2020zio}%
  \BibitemOpen
  \bibfield  {author} {\bibinfo {author} {\bibfnamefont {E.}~\bibnamefont
  {Di~Valentino}} \emph {et~al.},\ }\href {\doibase
  10.1016/j.astropartphys.2021.102605} {\bibfield  {journal} {\bibinfo
  {journal} {Astropart. Phys.}\ }\textbf {\bibinfo {volume} {131}},\ \bibinfo
  {pages} {102605} (\bibinfo {year} {2021}{\natexlab{a}})},\ \Eprint
  {http://arxiv.org/abs/2008.11284} {arXiv:2008.11284 [astro-ph.CO]}
  \BibitemShut {NoStop}%
\bibitem [{\citenamefont {Di~Valentino}\ \emph
  {et~al.}(2021{\natexlab{b}})\citenamefont {Di~Valentino}, \citenamefont
  {Mena}, \citenamefont {Pan}, \citenamefont {Visinelli}, \citenamefont {Yang},
  \citenamefont {Melchiorri}, \citenamefont {Mota}, \citenamefont {Riess},\
  and\ \citenamefont {Silk}}]{DiValentino:2021izs}%
  \BibitemOpen
  \bibfield  {author} {\bibinfo {author} {\bibfnamefont {E.}~\bibnamefont
  {Di~Valentino}}, \bibinfo {author} {\bibfnamefont {O.}~\bibnamefont {Mena}},
  \bibinfo {author} {\bibfnamefont {S.}~\bibnamefont {Pan}}, \bibinfo {author}
  {\bibfnamefont {L.}~\bibnamefont {Visinelli}}, \bibinfo {author}
  {\bibfnamefont {W.}~\bibnamefont {Yang}}, \bibinfo {author} {\bibfnamefont
  {A.}~\bibnamefont {Melchiorri}}, \bibinfo {author} {\bibfnamefont {D.~F.}\
  \bibnamefont {Mota}}, \bibinfo {author} {\bibfnamefont {A.~G.}\ \bibnamefont
  {Riess}}, \ and\ \bibinfo {author} {\bibfnamefont {J.}~\bibnamefont {Silk}},\
  }\href {\doibase 10.1088/1361-6382/ac086d} {\bibfield  {journal} {\bibinfo
  {journal} {Class. Quant. Grav.}\ }\textbf {\bibinfo {volume} {38}},\ \bibinfo
  {pages} {153001} (\bibinfo {year} {2021}{\natexlab{b}})},\ \Eprint
  {http://arxiv.org/abs/2103.01183} {arXiv:2103.01183 [astro-ph.CO]}
  \BibitemShut {NoStop}%
\bibitem [{\citenamefont {Perivolaropoulos}\ and\ \citenamefont
  {Skara}(2022)}]{Perivolaropoulos:2021jda}%
  \BibitemOpen
  \bibfield  {author} {\bibinfo {author} {\bibfnamefont {L.}~\bibnamefont
  {Perivolaropoulos}}\ and\ \bibinfo {author} {\bibfnamefont {F.}~\bibnamefont
  {Skara}},\ }\href {\doibase 10.1016/j.newar.2022.101659} {\bibfield
  {journal} {\bibinfo  {journal} {New Astron. Rev.}\ }\textbf {\bibinfo
  {volume} {95}},\ \bibinfo {pages} {101659} (\bibinfo {year} {2022})},\
  \Eprint {http://arxiv.org/abs/2105.05208} {arXiv:2105.05208 [astro-ph.CO]}
  \BibitemShut {NoStop}%
\bibitem [{\citenamefont {Abdalla}\ \emph {et~al.}(2022)\citenamefont {Abdalla}
  \emph {et~al.}}]{Abdalla:2022yfr}%
  \BibitemOpen
  \bibfield  {author} {\bibinfo {author} {\bibfnamefont {E.}~\bibnamefont
  {Abdalla}} \emph {et~al.},\ }\href {\doibase 10.1016/j.jheap.2022.04.002}
  {\bibfield  {journal} {\bibinfo  {journal} {JHEAp}\ }\textbf {\bibinfo
  {volume} {34}},\ \bibinfo {pages} {49} (\bibinfo {year} {2022})},\ \Eprint
  {http://arxiv.org/abs/2203.06142} {arXiv:2203.06142 [astro-ph.CO]}
  \BibitemShut {NoStop}%
\bibitem [{\citenamefont {Chevallier}\ and\ \citenamefont
  {Polarski}(2001)}]{Chevallier:2000qy}%
  \BibitemOpen
  \bibfield  {author} {\bibinfo {author} {\bibfnamefont {M.}~\bibnamefont
  {Chevallier}}\ and\ \bibinfo {author} {\bibfnamefont {D.}~\bibnamefont
  {Polarski}},\ }\href {\doibase 10.1142/S0218271801000822} {\bibfield
  {journal} {\bibinfo  {journal} {Int. J. Mod. Phys. D}\ }\textbf {\bibinfo
  {volume} {10}},\ \bibinfo {pages} {213} (\bibinfo {year} {2001})},\ \Eprint
  {http://arxiv.org/abs/gr-qc/0009008} {arXiv:gr-qc/0009008} \BibitemShut
  {NoStop}%
\bibitem [{\citenamefont {Linder}(2003)}]{Linder:2002et}%
  \BibitemOpen
  \bibfield  {author} {\bibinfo {author} {\bibfnamefont {E.~V.}\ \bibnamefont
  {Linder}},\ }\href {\doibase 10.1103/PhysRevLett.90.091301} {\bibfield
  {journal} {\bibinfo  {journal} {Phys. Rev. Lett.}\ }\textbf {\bibinfo
  {volume} {90}},\ \bibinfo {pages} {091301} (\bibinfo {year} {2003})},\
  \Eprint {http://arxiv.org/abs/astro-ph/0208512} {arXiv:astro-ph/0208512}
  \BibitemShut {NoStop}%
\bibitem [{\citenamefont {Ratra}\ and\ \citenamefont
  {Peebles}(1988)}]{Ratra:1987rm}%
  \BibitemOpen
  \bibfield  {author} {\bibinfo {author} {\bibfnamefont {B.}~\bibnamefont
  {Ratra}}\ and\ \bibinfo {author} {\bibfnamefont {P.~J.~E.}\ \bibnamefont
  {Peebles}},\ }\href {\doibase 10.1103/PhysRevD.37.3406} {\bibfield  {journal}
  {\bibinfo  {journal} {Phys. Rev. D}\ }\textbf {\bibinfo {volume} {37}},\
  \bibinfo {pages} {3406} (\bibinfo {year} {1988})}\BibitemShut {NoStop}%
\bibitem [{\citenamefont {Zlatev}\ \emph {et~al.}(1999)\citenamefont {Zlatev},
  \citenamefont {Wang},\ and\ \citenamefont {Steinhardt}}]{Zlatev:1998tr}%
  \BibitemOpen
  \bibfield  {author} {\bibinfo {author} {\bibfnamefont {I.}~\bibnamefont
  {Zlatev}}, \bibinfo {author} {\bibfnamefont {L.-M.}\ \bibnamefont {Wang}}, \
  and\ \bibinfo {author} {\bibfnamefont {P.~J.}\ \bibnamefont {Steinhardt}},\
  }\href {\doibase 10.1103/PhysRevLett.82.896} {\bibfield  {journal} {\bibinfo
  {journal} {Phys. Rev. Lett.}\ }\textbf {\bibinfo {volume} {82}},\ \bibinfo
  {pages} {896} (\bibinfo {year} {1999})},\ \Eprint
  {http://arxiv.org/abs/astro-ph/9807002} {arXiv:astro-ph/9807002} \BibitemShut
  {NoStop}%
\bibitem [{\citenamefont {Levi}\ \emph {et~al.}(2013)\citenamefont {Levi} \emph
  {et~al.}}]{DESI:2013agm}%
  \BibitemOpen
  \bibfield  {author} {\bibinfo {author} {\bibfnamefont {M.}~\bibnamefont
  {Levi}} \emph {et~al.} (\bibinfo {collaboration} {DESI}),\ }\href@noop {} {\
  (\bibinfo {year} {2013})},\ \Eprint {http://arxiv.org/abs/1308.0847}
  {arXiv:1308.0847 [astro-ph.CO]} \BibitemShut {NoStop}%
\bibitem [{\citenamefont {Aghamousa}\ \emph
  {et~al.}(2016{\natexlab{a}})\citenamefont {Aghamousa} \emph
  {et~al.}}]{DESI:2016fyo}%
  \BibitemOpen
  \bibfield  {author} {\bibinfo {author} {\bibfnamefont {A.}~\bibnamefont
  {Aghamousa}} \emph {et~al.} (\bibinfo {collaboration} {DESI}),\ }\href@noop
  {} {\  (\bibinfo {year} {2016}{\natexlab{a}})},\ \Eprint
  {http://arxiv.org/abs/1611.00036} {arXiv:1611.00036 [astro-ph.IM]}
  \BibitemShut {NoStop}%
\bibitem [{\citenamefont {Aghamousa}\ \emph
  {et~al.}(2016{\natexlab{b}})\citenamefont {Aghamousa} \emph
  {et~al.}}]{DESI:2016igz}%
  \BibitemOpen
  \bibfield  {author} {\bibinfo {author} {\bibfnamefont {A.}~\bibnamefont
  {Aghamousa}} \emph {et~al.} (\bibinfo {collaboration} {DESI}),\ }\href@noop
  {} {\  (\bibinfo {year} {2016}{\natexlab{b}})},\ \Eprint
  {http://arxiv.org/abs/1611.00037} {arXiv:1611.00037 [astro-ph.IM]}
  \BibitemShut {NoStop}%
\bibitem [{\citenamefont {Scolnic}\ \emph {et~al.}(2022)\citenamefont {Scolnic}
  \emph {et~al.}}]{Scolnic:2021amr}%
  \BibitemOpen
  \bibfield  {author} {\bibinfo {author} {\bibfnamefont {D.}~\bibnamefont
  {Scolnic}} \emph {et~al.},\ }\href {\doibase 10.3847/1538-4357/ac8b7a}
  {\bibfield  {journal} {\bibinfo  {journal} {Astrophys. J.}\ }\textbf
  {\bibinfo {volume} {938}},\ \bibinfo {pages} {113} (\bibinfo {year}
  {2022})},\ \Eprint {http://arxiv.org/abs/2112.03863} {arXiv:2112.03863
  [astro-ph.CO]} \BibitemShut {NoStop}%
\bibitem [{\citenamefont {Rubin}\ \emph {et~al.}(2023)\citenamefont {Rubin}
  \emph {et~al.}}]{Rubin:2023ovl}%
  \BibitemOpen
  \bibfield  {author} {\bibinfo {author} {\bibfnamefont {D.}~\bibnamefont
  {Rubin}} \emph {et~al.},\ }\href@noop {} {\  (\bibinfo {year} {2023})},\
  \Eprint {http://arxiv.org/abs/2311.12098} {arXiv:2311.12098 [astro-ph.CO]}
  \BibitemShut {NoStop}%
\bibitem [{\citenamefont {Abbott}\ \emph {et~al.}(2024)\citenamefont {Abbott}
  \emph {et~al.}}]{DES:2024jxu}%
  \BibitemOpen
  \bibfield  {author} {\bibinfo {author} {\bibfnamefont {T.~M.~C.}\
  \bibnamefont {Abbott}} \emph {et~al.} (\bibinfo {collaboration} {DES}),\
  }\href {\doibase 10.3847/2041-8213/ad6f9f} {\bibfield  {journal} {\bibinfo
  {journal} {Astrophys. J. Lett.}\ }\textbf {\bibinfo {volume} {973}},\
  \bibinfo {pages} {L14} (\bibinfo {year} {2024})},\ \Eprint
  {http://arxiv.org/abs/2401.02929} {arXiv:2401.02929 [astro-ph.CO]}
  \BibitemShut {NoStop}%
\bibitem [{\citenamefont {Abdul~Karim}\ \emph
  {et~al.}(2025{\natexlab{a}})\citenamefont {Abdul~Karim} \emph
  {et~al.}}]{DESI:2025zpo}%
  \BibitemOpen
  \bibfield  {author} {\bibinfo {author} {\bibfnamefont {M.}~\bibnamefont
  {Abdul~Karim}} \emph {et~al.} (\bibinfo {collaboration} {DESI}),\ }\href
  {\doibase 10.1103/2wwn-xjm5} {\bibfield  {journal} {\bibinfo  {journal}
  {Phys. Rev. D}\ } (\bibinfo {year} {2025}{\natexlab{a}}),\
  10.1103/2wwn-xjm5},\ \Eprint {http://arxiv.org/abs/2503.14739}
  {arXiv:2503.14739 [astro-ph.CO]} \BibitemShut {NoStop}%
\bibitem [{\citenamefont {Abdul~Karim}\ \emph
  {et~al.}(2025{\natexlab{b}})\citenamefont {Abdul~Karim} \emph
  {et~al.}}]{DESI:2025zgx}%
  \BibitemOpen
  \bibfield  {author} {\bibinfo {author} {\bibfnamefont {M.}~\bibnamefont
  {Abdul~Karim}} \emph {et~al.} (\bibinfo {collaboration} {DESI}),\ }\href@noop
  {} {\  (\bibinfo {year} {2025}{\natexlab{b}})},\ \Eprint
  {http://arxiv.org/abs/2503.14738} {arXiv:2503.14738 [astro-ph.CO]}
  \BibitemShut {NoStop}%
\bibitem [{\citenamefont {Lodha}\ \emph {et~al.}(2025)\citenamefont {Lodha}
  \emph {et~al.}}]{DESI:2025fii}%
  \BibitemOpen
  \bibfield  {author} {\bibinfo {author} {\bibfnamefont {K.}~\bibnamefont
  {Lodha}} \emph {et~al.} (\bibinfo {collaboration} {DESI}),\ }\href@noop {} {\
   (\bibinfo {year} {2025})},\ \Eprint {http://arxiv.org/abs/2503.14743}
  {arXiv:2503.14743 [astro-ph.CO]} \BibitemShut {NoStop}%
\bibitem [{\citenamefont {Yang}\ \emph
  {et~al.}(2025{\natexlab{a}})\citenamefont {Yang}, \citenamefont {Wang},
  \citenamefont {Li}, \citenamefont {Yuan}, \citenamefont {Ren}, \citenamefont
  {Saridakis},\ and\ \citenamefont {Cai}}]{Yang:2025kgc}%
  \BibitemOpen
  \bibfield  {author} {\bibinfo {author} {\bibfnamefont {Y.}~\bibnamefont
  {Yang}}, \bibinfo {author} {\bibfnamefont {Q.}~\bibnamefont {Wang}}, \bibinfo
  {author} {\bibfnamefont {C.}~\bibnamefont {Li}}, \bibinfo {author}
  {\bibfnamefont {P.}~\bibnamefont {Yuan}}, \bibinfo {author} {\bibfnamefont
  {X.}~\bibnamefont {Ren}}, \bibinfo {author} {\bibfnamefont {E.~N.}\
  \bibnamefont {Saridakis}}, \ and\ \bibinfo {author} {\bibfnamefont {Y.-F.}\
  \bibnamefont {Cai}},\ }\href {\doibase 10.1088/1475-7516/2025/08/050}
  {\bibfield  {journal} {\bibinfo  {journal} {JCAP}\ }\textbf {\bibinfo
  {volume} {08}},\ \bibinfo {pages} {050} (\bibinfo {year}
  {2025}{\natexlab{a}})},\ \Eprint {http://arxiv.org/abs/2501.18336}
  {arXiv:2501.18336 [astro-ph.CO]} \BibitemShut {NoStop}%
\bibitem [{\citenamefont {Yang}\ \emph {et~al.}(2024)\citenamefont {Yang},
  \citenamefont {Ren}, \citenamefont {Wang}, \citenamefont {Lu}, \citenamefont
  {Zhang}, \citenamefont {Cai},\ and\ \citenamefont
  {Saridakis}}]{Yang:2024kdo}%
  \BibitemOpen
  \bibfield  {author} {\bibinfo {author} {\bibfnamefont {Y.}~\bibnamefont
  {Yang}}, \bibinfo {author} {\bibfnamefont {X.}~\bibnamefont {Ren}}, \bibinfo
  {author} {\bibfnamefont {Q.}~\bibnamefont {Wang}}, \bibinfo {author}
  {\bibfnamefont {Z.}~\bibnamefont {Lu}}, \bibinfo {author} {\bibfnamefont
  {D.}~\bibnamefont {Zhang}}, \bibinfo {author} {\bibfnamefont {Y.-F.}\
  \bibnamefont {Cai}}, \ and\ \bibinfo {author} {\bibfnamefont {E.~N.}\
  \bibnamefont {Saridakis}},\ }\href {\doibase 10.1016/j.scib.2024.07.029}
  {\bibfield  {journal} {\bibinfo  {journal} {Sci. Bull.}\ }\textbf {\bibinfo
  {volume} {69}},\ \bibinfo {pages} {2698} (\bibinfo {year} {2024})},\ \Eprint
  {http://arxiv.org/abs/2404.19437} {arXiv:2404.19437 [astro-ph.CO]}
  \BibitemShut {NoStop}%
\bibitem [{\citenamefont {Yang}\ \emph
  {et~al.}(2025{\natexlab{b}})\citenamefont {Yang}, \citenamefont {Wang},
  \citenamefont {Ren}, \citenamefont {Saridakis},\ and\ \citenamefont
  {Cai}}]{Yang:2025mws}%
  \BibitemOpen
  \bibfield  {author} {\bibinfo {author} {\bibfnamefont {Y.}~\bibnamefont
  {Yang}}, \bibinfo {author} {\bibfnamefont {Q.}~\bibnamefont {Wang}}, \bibinfo
  {author} {\bibfnamefont {X.}~\bibnamefont {Ren}}, \bibinfo {author}
  {\bibfnamefont {E.~N.}\ \bibnamefont {Saridakis}}, \ and\ \bibinfo {author}
  {\bibfnamefont {Y.-F.}\ \bibnamefont {Cai}},\ }\href {\doibase
  10.3847/1538-4357/ade43f} {\bibfield  {journal} {\bibinfo  {journal}
  {Astrophys. J.}\ }\textbf {\bibinfo {volume} {988}},\ \bibinfo {pages} {123}
  (\bibinfo {year} {2025}{\natexlab{b}})},\ \Eprint
  {http://arxiv.org/abs/2504.06784} {arXiv:2504.06784 [astro-ph.CO]}
  \BibitemShut {NoStop}%
\bibitem [{\citenamefont {Strauss}\ \emph {et~al.}(2002)\citenamefont {Strauss}
  \emph {et~al.}}]{SDSS:2002jsq}%
  \BibitemOpen
  \bibfield  {author} {\bibinfo {author} {\bibfnamefont {M.~A.}\ \bibnamefont
  {Strauss}} \emph {et~al.} (\bibinfo {collaboration} {SDSS}),\ }\href
  {\doibase 10.1086/342343} {\bibfield  {journal} {\bibinfo  {journal} {Astron.
  J.}\ }\textbf {\bibinfo {volume} {124}},\ \bibinfo {pages} {1810} (\bibinfo
  {year} {2002})},\ \Eprint {http://arxiv.org/abs/astro-ph/0206225}
  {arXiv:astro-ph/0206225} \BibitemShut {NoStop}%
\bibitem [{\citenamefont {Eisenstein}\ \emph {et~al.}(2011)\citenamefont
  {Eisenstein} \emph {et~al.}}]{2011AJ....142...72E}%
  \BibitemOpen
  \bibfield  {author} {\bibinfo {author} {\bibfnamefont {D.~J.}\ \bibnamefont
  {Eisenstein}} \emph {et~al.} (\bibinfo {collaboration} {BOSS}),\ }\href
  {\doibase 10.1088/0004-6256/142/3/72} {\bibfield  {journal} {\bibinfo
  {journal} {Astron. J.}\ }\textbf {\bibinfo {volume} {142}},\ \bibinfo {pages}
  {72} (\bibinfo {year} {2011})},\ \Eprint {http://arxiv.org/abs/1101.1529}
  {arXiv:1101.1529 [astro-ph.IM]} \BibitemShut {NoStop}%
\bibitem [{\citenamefont {Choi}\ \emph {et~al.}(2021)\citenamefont {Choi},
  \citenamefont {Lin}, \citenamefont {Visinelli},\ and\ \citenamefont
  {Yanagida}}]{Choi:2021aze}%
  \BibitemOpen
  \bibfield  {author} {\bibinfo {author} {\bibfnamefont {G.}~\bibnamefont
  {Choi}}, \bibinfo {author} {\bibfnamefont {W.}~\bibnamefont {Lin}}, \bibinfo
  {author} {\bibfnamefont {L.}~\bibnamefont {Visinelli}}, \ and\ \bibinfo
  {author} {\bibfnamefont {T.~T.}\ \bibnamefont {Yanagida}},\ }\href {\doibase
  10.1103/PhysRevD.104.L101302} {\bibfield  {journal} {\bibinfo  {journal}
  {Phys. Rev. D}\ }\textbf {\bibinfo {volume} {104}},\ \bibinfo {pages}
  {L101302} (\bibinfo {year} {2021})},\ \Eprint
  {http://arxiv.org/abs/2106.12602} {arXiv:2106.12602 [hep-ph]} \BibitemShut
  {NoStop}%
\bibitem [{\citenamefont {Visinelli}\ and\ \citenamefont
  {Vagnozzi}(2019)}]{Visinelli:2018utg}%
  \BibitemOpen
  \bibfield  {author} {\bibinfo {author} {\bibfnamefont {L.}~\bibnamefont
  {Visinelli}}\ and\ \bibinfo {author} {\bibfnamefont {S.}~\bibnamefont
  {Vagnozzi}},\ }\href {\doibase 10.1103/PhysRevD.99.063517} {\bibfield
  {journal} {\bibinfo  {journal} {Phys. Rev. D}\ }\textbf {\bibinfo {volume}
  {99}},\ \bibinfo {pages} {063517} (\bibinfo {year} {2019})},\ \Eprint
  {http://arxiv.org/abs/1809.06382} {arXiv:1809.06382 [hep-ph]} \BibitemShut
  {NoStop}%
\bibitem [{\citenamefont {Reig}(2021)}]{Reig:2021ipa}%
  \BibitemOpen
  \bibfield  {author} {\bibinfo {author} {\bibfnamefont {M.}~\bibnamefont
  {Reig}},\ }\href {\doibase 10.1007/JHEP09(2021)207} {\bibfield  {journal}
  {\bibinfo  {journal} {JHEP}\ }\textbf {\bibinfo {volume} {09}},\ \bibinfo
  {pages} {207} (\bibinfo {year} {2021})},\ \Eprint
  {http://arxiv.org/abs/2104.09923} {arXiv:2104.09923 [hep-ph]} \BibitemShut
  {NoStop}%
\bibitem [{\citenamefont {Fukugita}\ and\ \citenamefont
  {Yanagida}(1994)}]{Fukugita:1994hq}%
  \BibitemOpen
  \bibfield  {author} {\bibinfo {author} {\bibfnamefont {M.}~\bibnamefont
  {Fukugita}}\ and\ \bibinfo {author} {\bibfnamefont {T.}~\bibnamefont
  {Yanagida}},\ }\href@noop {} {\  (\bibinfo {year} {1994})}\BibitemShut
  {NoStop}%
\bibitem [{\citenamefont {Frieman}\ \emph {et~al.}(1995)\citenamefont
  {Frieman}, \citenamefont {Hill}, \citenamefont {Stebbins},\ and\
  \citenamefont {Waga}}]{Frieman:1995pm}%
  \BibitemOpen
  \bibfield  {author} {\bibinfo {author} {\bibfnamefont {J.~A.}\ \bibnamefont
  {Frieman}}, \bibinfo {author} {\bibfnamefont {C.~T.}\ \bibnamefont {Hill}},
  \bibinfo {author} {\bibfnamefont {A.}~\bibnamefont {Stebbins}}, \ and\
  \bibinfo {author} {\bibfnamefont {I.}~\bibnamefont {Waga}},\ }\href {\doibase
  10.1103/PhysRevLett.75.2077} {\bibfield  {journal} {\bibinfo  {journal}
  {Phys. Rev. Lett.}\ }\textbf {\bibinfo {volume} {75}},\ \bibinfo {pages}
  {2077} (\bibinfo {year} {1995})},\ \Eprint
  {http://arxiv.org/abs/astro-ph/9505060} {arXiv:astro-ph/9505060} \BibitemShut
  {NoStop}%
\bibitem [{\citenamefont {Choi}(2000)}]{Choi:1999xn}%
  \BibitemOpen
  \bibfield  {author} {\bibinfo {author} {\bibfnamefont {K.}~\bibnamefont
  {Choi}},\ }\href {\doibase 10.1103/PhysRevD.62.043509} {\bibfield  {journal}
  {\bibinfo  {journal} {Phys. Rev. D}\ }\textbf {\bibinfo {volume} {62}},\
  \bibinfo {pages} {043509} (\bibinfo {year} {2000})},\ \Eprint
  {http://arxiv.org/abs/hep-ph/9902292} {arXiv:hep-ph/9902292} \BibitemShut
  {NoStop}%
\bibitem [{\citenamefont {Ng}\ and\ \citenamefont
  {Wiltshire}(2001)}]{Ng:2000di}%
  \BibitemOpen
  \bibfield  {author} {\bibinfo {author} {\bibfnamefont {S.~C.~C.}\
  \bibnamefont {Ng}}\ and\ \bibinfo {author} {\bibfnamefont {D.~L.}\
  \bibnamefont {Wiltshire}},\ }\href {\doibase 10.1103/PhysRevD.63.023503}
  {\bibfield  {journal} {\bibinfo  {journal} {Phys. Rev. D}\ }\textbf {\bibinfo
  {volume} {63}},\ \bibinfo {pages} {023503} (\bibinfo {year} {2001})},\
  \Eprint {http://arxiv.org/abs/astro-ph/0004138} {arXiv:astro-ph/0004138}
  \BibitemShut {NoStop}%
\bibitem [{\citenamefont {Nomura}\ \emph {et~al.}(2000)\citenamefont {Nomura},
  \citenamefont {Watari},\ and\ \citenamefont {Yanagida}}]{Nomura:2000yk}%
  \BibitemOpen
  \bibfield  {author} {\bibinfo {author} {\bibfnamefont {Y.}~\bibnamefont
  {Nomura}}, \bibinfo {author} {\bibfnamefont {T.}~\bibnamefont {Watari}}, \
  and\ \bibinfo {author} {\bibfnamefont {T.}~\bibnamefont {Yanagida}},\ }\href
  {\doibase 10.1016/S0370-2693(00)00605-5} {\bibfield  {journal} {\bibinfo
  {journal} {Phys. Lett. B}\ }\textbf {\bibinfo {volume} {484}},\ \bibinfo
  {pages} {103} (\bibinfo {year} {2000})},\ \Eprint
  {http://arxiv.org/abs/hep-ph/0004182} {arXiv:hep-ph/0004182} \BibitemShut
  {NoStop}%
\bibitem [{\citenamefont {Kawasaki}\ \emph {et~al.}(2001)\citenamefont
  {Kawasaki}, \citenamefont {Moroi},\ and\ \citenamefont
  {Takahashi}}]{Kawasaki:2001bq}%
  \BibitemOpen
  \bibfield  {author} {\bibinfo {author} {\bibfnamefont {M.}~\bibnamefont
  {Kawasaki}}, \bibinfo {author} {\bibfnamefont {T.}~\bibnamefont {Moroi}}, \
  and\ \bibinfo {author} {\bibfnamefont {T.}~\bibnamefont {Takahashi}},\ }\href
  {\doibase 10.1103/PhysRevD.64.083009} {\bibfield  {journal} {\bibinfo
  {journal} {Phys. Rev. D}\ }\textbf {\bibinfo {volume} {64}},\ \bibinfo
  {pages} {083009} (\bibinfo {year} {2001})},\ \Eprint
  {http://arxiv.org/abs/astro-ph/0105161} {arXiv:astro-ph/0105161} \BibitemShut
  {NoStop}%
\bibitem [{\citenamefont {Caldwell}\ and\ \citenamefont
  {Linder}(2005)}]{Caldwell:2005tm}%
  \BibitemOpen
  \bibfield  {author} {\bibinfo {author} {\bibfnamefont {R.~R.}\ \bibnamefont
  {Caldwell}}\ and\ \bibinfo {author} {\bibfnamefont {E.~V.}\ \bibnamefont
  {Linder}},\ }\href {\doibase 10.1103/PhysRevLett.95.141301} {\bibfield
  {journal} {\bibinfo  {journal} {Phys. Rev. Lett.}\ }\textbf {\bibinfo
  {volume} {95}},\ \bibinfo {pages} {141301} (\bibinfo {year} {2005})},\
  \Eprint {http://arxiv.org/abs/astro-ph/0505494} {arXiv:astro-ph/0505494}
  \BibitemShut {NoStop}%
\bibitem [{\citenamefont {Dutta}\ and\ \citenamefont
  {Sorbo}(2007)}]{Dutta:2006cf}%
  \BibitemOpen
  \bibfield  {author} {\bibinfo {author} {\bibfnamefont {K.}~\bibnamefont
  {Dutta}}\ and\ \bibinfo {author} {\bibfnamefont {L.}~\bibnamefont {Sorbo}},\
  }\href {\doibase 10.1103/PhysRevD.75.063514} {\bibfield  {journal} {\bibinfo
  {journal} {Phys. Rev. D}\ }\textbf {\bibinfo {volume} {75}},\ \bibinfo
  {pages} {063514} (\bibinfo {year} {2007})},\ \Eprint
  {http://arxiv.org/abs/astro-ph/0612457} {arXiv:astro-ph/0612457} \BibitemShut
  {NoStop}%
\bibitem [{\citenamefont {Riotto}\ and\ \citenamefont
  {Tkachev}(2000)}]{Riotto:2000kh}%
  \BibitemOpen
  \bibfield  {author} {\bibinfo {author} {\bibfnamefont {A.}~\bibnamefont
  {Riotto}}\ and\ \bibinfo {author} {\bibfnamefont {I.}~\bibnamefont
  {Tkachev}},\ }\href {\doibase 10.1016/S0370-2693(00)00660-2} {\bibfield
  {journal} {\bibinfo  {journal} {Phys. Lett. B}\ }\textbf {\bibinfo {volume}
  {484}},\ \bibinfo {pages} {177} (\bibinfo {year} {2000})},\ \Eprint
  {http://arxiv.org/abs/astro-ph/0003388} {arXiv:astro-ph/0003388} \BibitemShut
  {NoStop}%
\bibitem [{\citenamefont {Amendola}\ and\ \citenamefont
  {Barbieri}(2006)}]{Amendola:2005ad}%
  \BibitemOpen
  \bibfield  {author} {\bibinfo {author} {\bibfnamefont {L.}~\bibnamefont
  {Amendola}}\ and\ \bibinfo {author} {\bibfnamefont {R.}~\bibnamefont
  {Barbieri}},\ }\href {\doibase 10.1016/j.physletb.2006.08.069} {\bibfield
  {journal} {\bibinfo  {journal} {Phys. Lett. B}\ }\textbf {\bibinfo {volume}
  {642}},\ \bibinfo {pages} {192} (\bibinfo {year} {2006})},\ \Eprint
  {http://arxiv.org/abs/hep-ph/0509257} {arXiv:hep-ph/0509257} \BibitemShut
  {NoStop}%
\bibitem [{\citenamefont {de~Putter}\ and\ \citenamefont
  {Linder}(2008)}]{dePutter:2008wt}%
  \BibitemOpen
  \bibfield  {author} {\bibinfo {author} {\bibfnamefont {R.}~\bibnamefont
  {de~Putter}}\ and\ \bibinfo {author} {\bibfnamefont {E.~V.}\ \bibnamefont
  {Linder}},\ }\href {\doibase 10.1088/1475-7516/2008/10/042} {\bibfield
  {journal} {\bibinfo  {journal} {JCAP}\ }\textbf {\bibinfo {volume} {10}},\
  \bibinfo {pages} {042} (\bibinfo {year} {2008})},\ \Eprint
  {http://arxiv.org/abs/0808.0189} {arXiv:0808.0189 [astro-ph]} \BibitemShut
  {NoStop}%
\bibitem [{\citenamefont {Hlozek}\ \emph {et~al.}(2015)\citenamefont {Hlozek},
  \citenamefont {Grin}, \citenamefont {Marsh},\ and\ \citenamefont
  {Ferreira}}]{Hlozek:2014lca}%
  \BibitemOpen
  \bibfield  {author} {\bibinfo {author} {\bibfnamefont {R.}~\bibnamefont
  {Hlozek}}, \bibinfo {author} {\bibfnamefont {D.}~\bibnamefont {Grin}},
  \bibinfo {author} {\bibfnamefont {D.~J.~E.}\ \bibnamefont {Marsh}}, \ and\
  \bibinfo {author} {\bibfnamefont {P.~G.}\ \bibnamefont {Ferreira}},\ }\href
  {\doibase 10.1103/PhysRevD.91.103512} {\bibfield  {journal} {\bibinfo
  {journal} {Phys. Rev. D}\ }\textbf {\bibinfo {volume} {91}},\ \bibinfo
  {pages} {103512} (\bibinfo {year} {2015})},\ \Eprint
  {http://arxiv.org/abs/1410.2896} {arXiv:1410.2896 [astro-ph.CO]} \BibitemShut
  {NoStop}%
\bibitem [{\citenamefont {Lagu{\"e}}\ \emph {et~al.}(2022)\citenamefont
  {Lagu{\"e}}, \citenamefont {Bond}, \citenamefont {Hlo{\v{z}}ek},
  \citenamefont {Rogers}, \citenamefont {Marsh},\ and\ \citenamefont
  {Grin}}]{Lague:2021frh}%
  \BibitemOpen
  \bibfield  {author} {\bibinfo {author} {\bibfnamefont {A.}~\bibnamefont
  {Lagu{\"e}}}, \bibinfo {author} {\bibfnamefont {J.~R.}\ \bibnamefont {Bond}},
  \bibinfo {author} {\bibfnamefont {R.}~\bibnamefont {Hlo{\v{z}}ek}}, \bibinfo
  {author} {\bibfnamefont {K.~K.}\ \bibnamefont {Rogers}}, \bibinfo {author}
  {\bibfnamefont {D.~J.~E.}\ \bibnamefont {Marsh}}, \ and\ \bibinfo {author}
  {\bibfnamefont {D.}~\bibnamefont {Grin}},\ }\href {\doibase
  10.1088/1475-7516/2022/01/049} {\bibfield  {journal} {\bibinfo  {journal}
  {JCAP}\ }\textbf {\bibinfo {volume} {01}},\ \bibinfo {pages} {049} (\bibinfo
  {year} {2022})},\ \Eprint {http://arxiv.org/abs/2104.07802} {arXiv:2104.07802
  [astro-ph.CO]} \BibitemShut {NoStop}%
\bibitem [{\citenamefont {Rogers}\ \emph {et~al.}(2023)\citenamefont {Rogers},
  \citenamefont {Hlo{\v{z}}ek}, \citenamefont {Lagu{\"e}}, \citenamefont
  {Ivanov}, \citenamefont {Philcox}, \citenamefont {Cabass}, \citenamefont
  {Akitsu},\ and\ \citenamefont {Marsh}}]{Rogers:2023ezo}%
  \BibitemOpen
  \bibfield  {author} {\bibinfo {author} {\bibfnamefont {K.~K.}\ \bibnamefont
  {Rogers}}, \bibinfo {author} {\bibfnamefont {R.}~\bibnamefont
  {Hlo{\v{z}}ek}}, \bibinfo {author} {\bibfnamefont {A.}~\bibnamefont
  {Lagu{\"e}}}, \bibinfo {author} {\bibfnamefont {M.~M.}\ \bibnamefont
  {Ivanov}}, \bibinfo {author} {\bibfnamefont {O.~H.~E.}\ \bibnamefont
  {Philcox}}, \bibinfo {author} {\bibfnamefont {G.}~\bibnamefont {Cabass}},
  \bibinfo {author} {\bibfnamefont {K.}~\bibnamefont {Akitsu}}, \ and\ \bibinfo
  {author} {\bibfnamefont {D.~J.~E.}\ \bibnamefont {Marsh}},\ }\href {\doibase
  10.1088/1475-7516/2023/06/023} {\bibfield  {journal} {\bibinfo  {journal}
  {JCAP}\ }\textbf {\bibinfo {volume} {06}},\ \bibinfo {pages} {023} (\bibinfo
  {year} {2023})},\ \Eprint {http://arxiv.org/abs/2301.08361} {arXiv:2301.08361
  [astro-ph.CO]} \BibitemShut {NoStop}%
\bibitem [{\citenamefont {Berghaus}\ \emph {et~al.}(2024)\citenamefont
  {Berghaus}, \citenamefont {Kable},\ and\ \citenamefont
  {Miranda}}]{Berghaus:2024kra}%
  \BibitemOpen
  \bibfield  {author} {\bibinfo {author} {\bibfnamefont {K.~V.}\ \bibnamefont
  {Berghaus}}, \bibinfo {author} {\bibfnamefont {J.~A.}\ \bibnamefont {Kable}},
  \ and\ \bibinfo {author} {\bibfnamefont {V.}~\bibnamefont {Miranda}},\ }\href
  {\doibase 10.1103/PhysRevD.110.103524} {\bibfield  {journal} {\bibinfo
  {journal} {Phys. Rev. D}\ }\textbf {\bibinfo {volume} {110}},\ \bibinfo
  {pages} {103524} (\bibinfo {year} {2024})},\ \Eprint
  {http://arxiv.org/abs/2404.14341} {arXiv:2404.14341 [astro-ph.CO]}
  \BibitemShut {NoStop}%
\bibitem [{\citenamefont {Tada}\ and\ \citenamefont
  {Terada}(2024)}]{Tada:2024znt}%
  \BibitemOpen
  \bibfield  {author} {\bibinfo {author} {\bibfnamefont {Y.}~\bibnamefont
  {Tada}}\ and\ \bibinfo {author} {\bibfnamefont {T.}~\bibnamefont {Terada}},\
  }\href {\doibase 10.1103/PhysRevD.109.L121305} {\bibfield  {journal}
  {\bibinfo  {journal} {Phys. Rev. D}\ }\textbf {\bibinfo {volume} {109}},\
  \bibinfo {pages} {L121305} (\bibinfo {year} {2024})},\ \Eprint
  {http://arxiv.org/abs/2404.05722} {arXiv:2404.05722 [astro-ph.CO]}
  \BibitemShut {NoStop}%
\bibitem [{\citenamefont {Notari}\ \emph {et~al.}(2024)\citenamefont {Notari},
  \citenamefont {Redi},\ and\ \citenamefont {Tesi}}]{Notari:2024rti}%
  \BibitemOpen
  \bibfield  {author} {\bibinfo {author} {\bibfnamefont {A.}~\bibnamefont
  {Notari}}, \bibinfo {author} {\bibfnamefont {M.}~\bibnamefont {Redi}}, \ and\
  \bibinfo {author} {\bibfnamefont {A.}~\bibnamefont {Tesi}},\ }\href {\doibase
  10.1088/1475-7516/2024/11/025} {\bibfield  {journal} {\bibinfo  {journal}
  {JCAP}\ }\textbf {\bibinfo {volume} {11}},\ \bibinfo {pages} {025} (\bibinfo
  {year} {2024})},\ \Eprint {http://arxiv.org/abs/2406.08459} {arXiv:2406.08459
  [astro-ph.CO]} \BibitemShut {NoStop}%
\bibitem [{\citenamefont {Luu}\ \emph {et~al.}(2025)\citenamefont {Luu},
  \citenamefont {Qiu},\ and\ \citenamefont {Tye}}]{Luu:2025fgw}%
  \BibitemOpen
  \bibfield  {author} {\bibinfo {author} {\bibfnamefont {H.~N.}\ \bibnamefont
  {Luu}}, \bibinfo {author} {\bibfnamefont {Y.-C.}\ \bibnamefont {Qiu}}, \ and\
  \bibinfo {author} {\bibfnamefont {S.~H.~H.}\ \bibnamefont {Tye}},\ }\href
  {\doibase 10.1103/3mpg-24d2} {\bibfield  {journal} {\bibinfo  {journal}
  {Phys. Rev. D}\ }\textbf {\bibinfo {volume} {112}},\ \bibinfo {pages}
  {023524} (\bibinfo {year} {2025})},\ \Eprint
  {http://arxiv.org/abs/2503.18120} {arXiv:2503.18120 [hep-ph]} \BibitemShut
  {NoStop}%
\bibitem [{\citenamefont {Shlivko}\ and\ \citenamefont
  {Steinhardt}(2024)}]{Shlivko:2024llw}%
  \BibitemOpen
  \bibfield  {author} {\bibinfo {author} {\bibfnamefont {D.}~\bibnamefont
  {Shlivko}}\ and\ \bibinfo {author} {\bibfnamefont {P.~J.}\ \bibnamefont
  {Steinhardt}},\ }\href {\doibase 10.1016/j.physletb.2024.138826} {\bibfield
  {journal} {\bibinfo  {journal} {Phys. Lett. B}\ }\textbf {\bibinfo {volume}
  {855}},\ \bibinfo {pages} {138826} (\bibinfo {year} {2024})},\ \Eprint
  {http://arxiv.org/abs/2405.03933} {arXiv:2405.03933 [astro-ph.CO]}
  \BibitemShut {NoStop}%
\bibitem [{\citenamefont {Ure{\~n}a-L{\'o}pez}\ \emph
  {et~al.}(2025)\citenamefont {Ure{\~n}a-L{\'o}pez} \emph
  {et~al.}}]{Urena-Lopez:2025rad}%
  \BibitemOpen
  \bibfield  {author} {\bibinfo {author} {\bibfnamefont {L.~A.}\ \bibnamefont
  {Ure{\~n}a-L{\'o}pez}} \emph {et~al.},\ }\href@noop {} {\  (\bibinfo {year}
  {2025})},\ \Eprint {http://arxiv.org/abs/2503.20178} {arXiv:2503.20178
  [astro-ph.CO]} \BibitemShut {NoStop}%
\bibitem [{\citenamefont {Silva}\ \emph {et~al.}(2025)\citenamefont {Silva},
  \citenamefont {Sabogal}, \citenamefont {Scherer}, \citenamefont {Nunes},
  \citenamefont {Di~Valentino},\ and\ \citenamefont {Kumar}}]{Silva:2025hxw}%
  \BibitemOpen
  \bibfield  {author} {\bibinfo {author} {\bibfnamefont {E.}~\bibnamefont
  {Silva}}, \bibinfo {author} {\bibfnamefont {M.~A.}\ \bibnamefont {Sabogal}},
  \bibinfo {author} {\bibfnamefont {M.}~\bibnamefont {Scherer}}, \bibinfo
  {author} {\bibfnamefont {R.~C.}\ \bibnamefont {Nunes}}, \bibinfo {author}
  {\bibfnamefont {E.}~\bibnamefont {Di~Valentino}}, \ and\ \bibinfo {author}
  {\bibfnamefont {S.}~\bibnamefont {Kumar}},\ }\href {\doibase
  10.1103/qqc6-76z4} {\bibfield  {journal} {\bibinfo  {journal} {Phys. Rev. D}\
  }\textbf {\bibinfo {volume} {111}},\ \bibinfo {pages} {123511} (\bibinfo
  {year} {2025})},\ \Eprint {http://arxiv.org/abs/2503.23225} {arXiv:2503.23225
  [astro-ph.CO]} \BibitemShut {NoStop}%
\bibitem [{\citenamefont {Gialamas}\ \emph {et~al.}(2025)\citenamefont
  {Gialamas}, \citenamefont {H{\"u}tsi}, \citenamefont {Kannike}, \citenamefont
  {Racioppi}, \citenamefont {Raidal}, \citenamefont {Vasar},\ and\
  \citenamefont {Veerm{\"a}e}}]{Gialamas:2024lyw}%
  \BibitemOpen
  \bibfield  {author} {\bibinfo {author} {\bibfnamefont {I.~D.}\ \bibnamefont
  {Gialamas}}, \bibinfo {author} {\bibfnamefont {G.}~\bibnamefont {H{\"u}tsi}},
  \bibinfo {author} {\bibfnamefont {K.}~\bibnamefont {Kannike}}, \bibinfo
  {author} {\bibfnamefont {A.}~\bibnamefont {Racioppi}}, \bibinfo {author}
  {\bibfnamefont {M.}~\bibnamefont {Raidal}}, \bibinfo {author} {\bibfnamefont
  {M.}~\bibnamefont {Vasar}}, \ and\ \bibinfo {author} {\bibfnamefont
  {H.}~\bibnamefont {Veerm{\"a}e}},\ }\href {\doibase
  10.1103/PhysRevD.111.043540} {\bibfield  {journal} {\bibinfo  {journal}
  {Phys. Rev. D}\ }\textbf {\bibinfo {volume} {111}},\ \bibinfo {pages}
  {043540} (\bibinfo {year} {2025})},\ \Eprint
  {http://arxiv.org/abs/2406.07533} {arXiv:2406.07533 [astro-ph.CO]}
  \BibitemShut {NoStop}%
\bibitem [{\citenamefont {Banerjee}\ \emph {et~al.}(2021)\citenamefont
  {Banerjee}, \citenamefont {Cai}, \citenamefont {Heisenberg}, \citenamefont
  {Colg{\'a}in}, \citenamefont {Sheikh-Jabbari},\ and\ \citenamefont
  {Yang}}]{Banerjee:2020xcn}%
  \BibitemOpen
  \bibfield  {author} {\bibinfo {author} {\bibfnamefont {A.}~\bibnamefont
  {Banerjee}}, \bibinfo {author} {\bibfnamefont {H.}~\bibnamefont {Cai}},
  \bibinfo {author} {\bibfnamefont {L.}~\bibnamefont {Heisenberg}}, \bibinfo
  {author} {\bibfnamefont {E.~{\'O}.}\ \bibnamefont {Colg{\'a}in}}, \bibinfo
  {author} {\bibfnamefont {M.~M.}\ \bibnamefont {Sheikh-Jabbari}}, \ and\
  \bibinfo {author} {\bibfnamefont {T.}~\bibnamefont {Yang}},\ }\href {\doibase
  10.1103/PhysRevD.103.L081305} {\bibfield  {journal} {\bibinfo  {journal}
  {Phys. Rev. D}\ }\textbf {\bibinfo {volume} {103}},\ \bibinfo {pages}
  {L081305} (\bibinfo {year} {2021})},\ \Eprint
  {http://arxiv.org/abs/2006.00244} {arXiv:2006.00244 [astro-ph.CO]}
  \BibitemShut {NoStop}%
\bibitem [{\citenamefont {Lee}\ \emph {et~al.}(2022)\citenamefont {Lee},
  \citenamefont {Lee}, \citenamefont {Colg{\'a}in}, \citenamefont
  {Sheikh-Jabbari},\ and\ \citenamefont {Thakur}}]{Lee:2022cyh}%
  \BibitemOpen
  \bibfield  {author} {\bibinfo {author} {\bibfnamefont {B.-H.}\ \bibnamefont
  {Lee}}, \bibinfo {author} {\bibfnamefont {W.}~\bibnamefont {Lee}}, \bibinfo
  {author} {\bibfnamefont {E.~{\'O}.}\ \bibnamefont {Colg{\'a}in}}, \bibinfo
  {author} {\bibfnamefont {M.~M.}\ \bibnamefont {Sheikh-Jabbari}}, \ and\
  \bibinfo {author} {\bibfnamefont {S.}~\bibnamefont {Thakur}},\ }\href
  {\doibase 10.1088/1475-7516/2022/04/004} {\bibfield  {journal} {\bibinfo
  {journal} {JCAP}\ }\textbf {\bibinfo {volume} {04}},\ \bibinfo {pages} {004}
  (\bibinfo {year} {2022})},\ \Eprint {http://arxiv.org/abs/2202.03906}
  {arXiv:2202.03906 [astro-ph.CO]} \BibitemShut {NoStop}%
\bibitem [{\citenamefont {Colg{\'a}in}\ \emph {et~al.}(2025)\citenamefont
  {Colg{\'a}in}, \citenamefont {Pourojaghi}, \citenamefont {Sheikh-Jabbari},\
  and\ \citenamefont {Yin}}]{Colgain:2025nzf}%
  \BibitemOpen
  \bibfield  {author} {\bibinfo {author} {\bibfnamefont {E.~{\'O}.}\
  \bibnamefont {Colg{\'a}in}}, \bibinfo {author} {\bibfnamefont
  {S.}~\bibnamefont {Pourojaghi}}, \bibinfo {author} {\bibfnamefont {M.~M.}\
  \bibnamefont {Sheikh-Jabbari}}, \ and\ \bibinfo {author} {\bibfnamefont
  {L.}~\bibnamefont {Yin}},\ }\href@noop {} {\  (\bibinfo {year} {2025})},\
  \Eprint {http://arxiv.org/abs/2504.04417} {arXiv:2504.04417 [astro-ph.CO]}
  \BibitemShut {NoStop}%
\bibitem [{\citenamefont {Girmohanta}\ \emph {et~al.}(2023)\citenamefont
  {Girmohanta}, \citenamefont {Qiu}, \citenamefont {Wang},\ and\ \citenamefont
  {Yanagida}}]{Girmohanta:2023ghm}%
  \BibitemOpen
  \bibfield  {author} {\bibinfo {author} {\bibfnamefont {S.}~\bibnamefont
  {Girmohanta}}, \bibinfo {author} {\bibfnamefont {Y.-C.}\ \bibnamefont {Qiu}},
  \bibinfo {author} {\bibfnamefont {J.-W.}\ \bibnamefont {Wang}}, \ and\
  \bibinfo {author} {\bibfnamefont {T.~T.}\ \bibnamefont {Yanagida}},\ }\href
  {\doibase 10.1103/PhysRevD.108.015028} {\bibfield  {journal} {\bibinfo
  {journal} {Phys. Rev. D}\ }\textbf {\bibinfo {volume} {108}},\ \bibinfo
  {pages} {015028} (\bibinfo {year} {2023})},\ \Eprint
  {http://arxiv.org/abs/2303.02852} {arXiv:2303.02852 [hep-ph]} \BibitemShut
  {NoStop}%
\bibitem [{\citenamefont {Minami}\ and\ \citenamefont
  {Komatsu}(2020)}]{Minami:2020odp}%
  \BibitemOpen
  \bibfield  {author} {\bibinfo {author} {\bibfnamefont {Y.}~\bibnamefont
  {Minami}}\ and\ \bibinfo {author} {\bibfnamefont {E.}~\bibnamefont
  {Komatsu}},\ }\href {\doibase 10.1103/PhysRevLett.125.221301} {\bibfield
  {journal} {\bibinfo  {journal} {Phys. Rev. Lett.}\ }\textbf {\bibinfo
  {volume} {125}},\ \bibinfo {pages} {221301} (\bibinfo {year} {2020})},\
  \Eprint {http://arxiv.org/abs/2011.11254} {arXiv:2011.11254 [astro-ph.CO]}
  \BibitemShut {NoStop}%
\bibitem [{\citenamefont {Diego-Palazuelos}\ \emph {et~al.}(2022)\citenamefont
  {Diego-Palazuelos} \emph {et~al.}}]{Diego-Palazuelos:2022dsq}%
  \BibitemOpen
  \bibfield  {author} {\bibinfo {author} {\bibfnamefont {P.}~\bibnamefont
  {Diego-Palazuelos}} \emph {et~al.},\ }\href {\doibase
  10.1103/PhysRevLett.128.091302} {\bibfield  {journal} {\bibinfo  {journal}
  {Phys. Rev. Lett.}\ }\textbf {\bibinfo {volume} {128}},\ \bibinfo {pages}
  {091302} (\bibinfo {year} {2022})},\ \Eprint
  {http://arxiv.org/abs/2201.07682} {arXiv:2201.07682 [astro-ph.CO]}
  \BibitemShut {NoStop}%
\bibitem [{\citenamefont {Eskilt}(2022)}]{Eskilt:2022wav}%
  \BibitemOpen
  \bibfield  {author} {\bibinfo {author} {\bibfnamefont {J.~R.}\ \bibnamefont
  {Eskilt}},\ }\href {\doibase 10.1051/0004-6361/202243269} {\bibfield
  {journal} {\bibinfo  {journal} {Astron. Astrophys.}\ }\textbf {\bibinfo
  {volume} {662}},\ \bibinfo {pages} {A10} (\bibinfo {year} {2022})},\ \Eprint
  {http://arxiv.org/abs/2201.13347} {arXiv:2201.13347 [astro-ph.CO]}
  \BibitemShut {NoStop}%
\bibitem [{\citenamefont {Eskilt}\ and\ \citenamefont
  {Komatsu}(2022)}]{Eskilt:2022cff}%
  \BibitemOpen
  \bibfield  {author} {\bibinfo {author} {\bibfnamefont {J.~R.}\ \bibnamefont
  {Eskilt}}\ and\ \bibinfo {author} {\bibfnamefont {E.}~\bibnamefont
  {Komatsu}},\ }\href {\doibase 10.1103/PhysRevD.106.063503} {\bibfield
  {journal} {\bibinfo  {journal} {Phys. Rev. D}\ }\textbf {\bibinfo {volume}
  {106}},\ \bibinfo {pages} {063503} (\bibinfo {year} {2022})},\ \Eprint
  {http://arxiv.org/abs/2205.13962} {arXiv:2205.13962 [astro-ph.CO]}
  \BibitemShut {NoStop}%
\bibitem [{\citenamefont {Fujita}\ \emph {et~al.}(2021)\citenamefont {Fujita},
  \citenamefont {Murai}, \citenamefont {Nakatsuka},\ and\ \citenamefont
  {Tsujikawa}}]{Fujita:2020ecn}%
  \BibitemOpen
  \bibfield  {author} {\bibinfo {author} {\bibfnamefont {T.}~\bibnamefont
  {Fujita}}, \bibinfo {author} {\bibfnamefont {K.}~\bibnamefont {Murai}},
  \bibinfo {author} {\bibfnamefont {H.}~\bibnamefont {Nakatsuka}}, \ and\
  \bibinfo {author} {\bibfnamefont {S.}~\bibnamefont {Tsujikawa}},\ }\href
  {\doibase 10.1103/PhysRevD.103.043509} {\bibfield  {journal} {\bibinfo
  {journal} {Phys. Rev. D}\ }\textbf {\bibinfo {volume} {103}},\ \bibinfo
  {pages} {043509} (\bibinfo {year} {2021})},\ \Eprint
  {http://arxiv.org/abs/2011.11894} {arXiv:2011.11894 [astro-ph.CO]}
  \BibitemShut {NoStop}%
\bibitem [{\citenamefont {Takahashi}\ and\ \citenamefont
  {Yin}(2021)}]{Takahashi:2020tqv}%
  \BibitemOpen
  \bibfield  {author} {\bibinfo {author} {\bibfnamefont {F.}~\bibnamefont
  {Takahashi}}\ and\ \bibinfo {author} {\bibfnamefont {W.}~\bibnamefont
  {Yin}},\ }\href {\doibase 10.1088/1475-7516/2021/04/007} {\bibfield
  {journal} {\bibinfo  {journal} {JCAP}\ }\textbf {\bibinfo {volume} {04}},\
  \bibinfo {pages} {007} (\bibinfo {year} {2021})},\ \Eprint
  {http://arxiv.org/abs/2012.11576} {arXiv:2012.11576 [hep-ph]} \BibitemShut
  {NoStop}%
\bibitem [{\citenamefont {Fung}\ \emph {et~al.}(2021)\citenamefont {Fung},
  \citenamefont {Li}, \citenamefont {Liu}, \citenamefont {Luu}, \citenamefont
  {Qiu},\ and\ \citenamefont {Tye}}]{Fung:2021wbz}%
  \BibitemOpen
  \bibfield  {author} {\bibinfo {author} {\bibfnamefont {L.~W.~H.}\
  \bibnamefont {Fung}}, \bibinfo {author} {\bibfnamefont {L.}~\bibnamefont
  {Li}}, \bibinfo {author} {\bibfnamefont {T.}~\bibnamefont {Liu}}, \bibinfo
  {author} {\bibfnamefont {H.~N.}\ \bibnamefont {Luu}}, \bibinfo {author}
  {\bibfnamefont {Y.-C.}\ \bibnamefont {Qiu}}, \ and\ \bibinfo {author}
  {\bibfnamefont {S.~H.~H.}\ \bibnamefont {Tye}},\ }\href {\doibase
  10.1088/1475-7516/2021/08/057} {\bibfield  {journal} {\bibinfo  {journal}
  {JCAP}\ }\textbf {\bibinfo {volume} {08}},\ \bibinfo {pages} {057} (\bibinfo
  {year} {2021})},\ \Eprint {http://arxiv.org/abs/2102.11257} {arXiv:2102.11257
  [hep-ph]} \BibitemShut {NoStop}%
\bibitem [{\citenamefont {Nakagawa}\ \emph {et~al.}(2021)\citenamefont
  {Nakagawa}, \citenamefont {Takahashi},\ and\ \citenamefont
  {Yamada}}]{Nakagawa:2021nme}%
  \BibitemOpen
  \bibfield  {author} {\bibinfo {author} {\bibfnamefont {S.}~\bibnamefont
  {Nakagawa}}, \bibinfo {author} {\bibfnamefont {F.}~\bibnamefont {Takahashi}},
  \ and\ \bibinfo {author} {\bibfnamefont {M.}~\bibnamefont {Yamada}},\ }\href
  {\doibase 10.1103/PhysRevLett.127.181103} {\bibfield  {journal} {\bibinfo
  {journal} {Phys. Rev. Lett.}\ }\textbf {\bibinfo {volume} {127}},\ \bibinfo
  {pages} {181103} (\bibinfo {year} {2021})},\ \Eprint
  {http://arxiv.org/abs/2103.08153} {arXiv:2103.08153 [hep-ph]} \BibitemShut
  {NoStop}%
\bibitem [{\citenamefont {Jain}\ \emph {et~al.}(2021)\citenamefont {Jain},
  \citenamefont {Long},\ and\ \citenamefont {Amin}}]{Jain:2021shf}%
  \BibitemOpen
  \bibfield  {author} {\bibinfo {author} {\bibfnamefont {M.}~\bibnamefont
  {Jain}}, \bibinfo {author} {\bibfnamefont {A.~J.}\ \bibnamefont {Long}}, \
  and\ \bibinfo {author} {\bibfnamefont {M.~A.}\ \bibnamefont {Amin}},\ }\href
  {\doibase 10.1088/1475-7516/2021/05/055} {\bibfield  {journal} {\bibinfo
  {journal} {JCAP}\ }\textbf {\bibinfo {volume} {05}},\ \bibinfo {pages} {055}
  (\bibinfo {year} {2021})},\ \Eprint {http://arxiv.org/abs/2103.10962}
  {arXiv:2103.10962 [astro-ph.CO]} \BibitemShut {NoStop}%
\bibitem [{\citenamefont {Murai}\ \emph {et~al.}(2023)\citenamefont {Murai},
  \citenamefont {Naokawa}, \citenamefont {Namikawa},\ and\ \citenamefont
  {Komatsu}}]{Murai:2022zur}%
  \BibitemOpen
  \bibfield  {author} {\bibinfo {author} {\bibfnamefont {K.}~\bibnamefont
  {Murai}}, \bibinfo {author} {\bibfnamefont {F.}~\bibnamefont {Naokawa}},
  \bibinfo {author} {\bibfnamefont {T.}~\bibnamefont {Namikawa}}, \ and\
  \bibinfo {author} {\bibfnamefont {E.}~\bibnamefont {Komatsu}},\ }\href
  {\doibase 10.1103/PhysRevD.107.L041302} {\bibfield  {journal} {\bibinfo
  {journal} {Phys. Rev. D}\ }\textbf {\bibinfo {volume} {107}},\ \bibinfo
  {pages} {L041302} (\bibinfo {year} {2023})},\ \Eprint
  {http://arxiv.org/abs/2209.07804} {arXiv:2209.07804 [astro-ph.CO]}
  \BibitemShut {NoStop}%
\bibitem [{\citenamefont {Eskilt}\ \emph {et~al.}(2023)\citenamefont {Eskilt},
  \citenamefont {Herold}, \citenamefont {Komatsu}, \citenamefont {Murai},
  \citenamefont {Namikawa},\ and\ \citenamefont {Naokawa}}]{Eskilt:2023nxm}%
  \BibitemOpen
  \bibfield  {author} {\bibinfo {author} {\bibfnamefont {J.~R.}\ \bibnamefont
  {Eskilt}}, \bibinfo {author} {\bibfnamefont {L.}~\bibnamefont {Herold}},
  \bibinfo {author} {\bibfnamefont {E.}~\bibnamefont {Komatsu}}, \bibinfo
  {author} {\bibfnamefont {K.}~\bibnamefont {Murai}}, \bibinfo {author}
  {\bibfnamefont {T.}~\bibnamefont {Namikawa}}, \ and\ \bibinfo {author}
  {\bibfnamefont {F.}~\bibnamefont {Naokawa}},\ }\href {\doibase
  10.1103/PhysRevLett.131.121001} {\bibfield  {journal} {\bibinfo  {journal}
  {Phys. Rev. Lett.}\ }\textbf {\bibinfo {volume} {131}},\ \bibinfo {pages}
  {121001} (\bibinfo {year} {2023})},\ \Eprint
  {http://arxiv.org/abs/2303.15369} {arXiv:2303.15369 [astro-ph.CO]}
  \BibitemShut {NoStop}%
\bibitem [{\citenamefont {Nakagawa}\ \emph {et~al.}(2025)\citenamefont
  {Nakagawa}, \citenamefont {Nakai}, \citenamefont {Qiu},\ and\ \citenamefont
  {Yamada}}]{Nakagawa:2025ejs}%
  \BibitemOpen
  \bibfield  {author} {\bibinfo {author} {\bibfnamefont {S.}~\bibnamefont
  {Nakagawa}}, \bibinfo {author} {\bibfnamefont {Y.}~\bibnamefont {Nakai}},
  \bibinfo {author} {\bibfnamefont {Y.-C.}\ \bibnamefont {Qiu}}, \ and\
  \bibinfo {author} {\bibfnamefont {M.}~\bibnamefont {Yamada}},\ }\href
  {\doibase 10.1016/j.physletb.2025.139774} {\bibfield  {journal} {\bibinfo
  {journal} {Phys. Lett. B}\ }\textbf {\bibinfo {volume} {868}},\ \bibinfo
  {pages} {139774} (\bibinfo {year} {2025})},\ \Eprint
  {http://arxiv.org/abs/2503.18924} {arXiv:2503.18924 [astro-ph.CO]}
  \BibitemShut {NoStop}%
\bibitem [{\citenamefont {Lee}\ \emph {et~al.}(2025)\citenamefont {Lee},
  \citenamefont {Murai}, \citenamefont {Takahashi},\ and\ \citenamefont
  {Yin}}]{Lee:2025yvn}%
  \BibitemOpen
  \bibfield  {author} {\bibinfo {author} {\bibfnamefont {J.}~\bibnamefont
  {Lee}}, \bibinfo {author} {\bibfnamefont {K.}~\bibnamefont {Murai}}, \bibinfo
  {author} {\bibfnamefont {F.}~\bibnamefont {Takahashi}}, \ and\ \bibinfo
  {author} {\bibfnamefont {W.}~\bibnamefont {Yin}},\ }\href@noop {} {\
  (\bibinfo {year} {2025})},\ \Eprint {http://arxiv.org/abs/2503.18417}
  {arXiv:2503.18417 [hep-ph]} \BibitemShut {NoStop}%
\bibitem [{\citenamefont {Lemos}\ and\ \citenamefont
  {Lewis}(2023)}]{Lemos:2023xhs}%
  \BibitemOpen
  \bibfield  {author} {\bibinfo {author} {\bibfnamefont {P.}~\bibnamefont
  {Lemos}}\ and\ \bibinfo {author} {\bibfnamefont {A.}~\bibnamefont {Lewis}},\
  }\href {\doibase 10.1103/PhysRevD.107.103505} {\bibfield  {journal} {\bibinfo
   {journal} {Phys. Rev. D}\ }\textbf {\bibinfo {volume} {107}},\ \bibinfo
  {pages} {103505} (\bibinfo {year} {2023})},\ \Eprint
  {http://arxiv.org/abs/2302.12911} {arXiv:2302.12911 [astro-ph.CO]}
  \BibitemShut {NoStop}%
\bibitem [{\citenamefont {Prince}\ and\ \citenamefont
  {Dunkley}(2019)}]{Prince:2019hse}%
  \BibitemOpen
  \bibfield  {author} {\bibinfo {author} {\bibfnamefont {H.}~\bibnamefont
  {Prince}}\ and\ \bibinfo {author} {\bibfnamefont {J.}~\bibnamefont
  {Dunkley}},\ }\href {\doibase 10.1103/PhysRevD.100.083502} {\bibfield
  {journal} {\bibinfo  {journal} {Phys. Rev. D}\ }\textbf {\bibinfo {volume}
  {100}},\ \bibinfo {pages} {083502} (\bibinfo {year} {2019})},\ \Eprint
  {http://arxiv.org/abs/1909.05869} {arXiv:1909.05869 [astro-ph.CO]}
  \BibitemShut {NoStop}%
\bibitem [{\citenamefont {Hu}\ \emph {et~al.}(2001)\citenamefont {Hu},
  \citenamefont {Fukugita}, \citenamefont {Zaldarriaga},\ and\ \citenamefont
  {Tegmark}}]{Hu:2000ti}%
  \BibitemOpen
  \bibfield  {author} {\bibinfo {author} {\bibfnamefont {W.}~\bibnamefont
  {Hu}}, \bibinfo {author} {\bibfnamefont {M.}~\bibnamefont {Fukugita}},
  \bibinfo {author} {\bibfnamefont {M.}~\bibnamefont {Zaldarriaga}}, \ and\
  \bibinfo {author} {\bibfnamefont {M.}~\bibnamefont {Tegmark}},\ }\href
  {\doibase 10.1086/319449} {\bibfield  {journal} {\bibinfo  {journal}
  {Astrophys. J.}\ }\textbf {\bibinfo {volume} {549}},\ \bibinfo {pages} {669}
  (\bibinfo {year} {2001})},\ \Eprint {http://arxiv.org/abs/astro-ph/0006436}
  {arXiv:astro-ph/0006436} \BibitemShut {NoStop}%
\bibitem [{\citenamefont {Lin}\ \emph {et~al.}(2021)\citenamefont {Lin},
  \citenamefont {Chen},\ and\ \citenamefont {Mack}}]{Lin:2021sfs}%
  \BibitemOpen
  \bibfield  {author} {\bibinfo {author} {\bibfnamefont {W.}~\bibnamefont
  {Lin}}, \bibinfo {author} {\bibfnamefont {X.}~\bibnamefont {Chen}}, \ and\
  \bibinfo {author} {\bibfnamefont {K.~J.}\ \bibnamefont {Mack}},\ }\href
  {\doibase 10.3847/1538-4357/ac12cf} {\bibfield  {journal} {\bibinfo
  {journal} {Astrophys. J.}\ }\textbf {\bibinfo {volume} {920}},\ \bibinfo
  {pages} {159} (\bibinfo {year} {2021})},\ \Eprint
  {http://arxiv.org/abs/2102.05701} {arXiv:2102.05701 [astro-ph.CO]}
  \BibitemShut {NoStop}%
\bibitem [{\citenamefont {Hu}\ and\ \citenamefont
  {Sugiyama}(1996)}]{Hu:1995en}%
  \BibitemOpen
  \bibfield  {author} {\bibinfo {author} {\bibfnamefont {W.}~\bibnamefont
  {Hu}}\ and\ \bibinfo {author} {\bibfnamefont {N.}~\bibnamefont {Sugiyama}},\
  }\href {\doibase 10.1086/177989} {\bibfield  {journal} {\bibinfo  {journal}
  {Astrophys. J.}\ }\textbf {\bibinfo {volume} {471}},\ \bibinfo {pages} {542}
  (\bibinfo {year} {1996})},\ \Eprint {http://arxiv.org/abs/astro-ph/9510117}
  {arXiv:astro-ph/9510117} \BibitemShut {NoStop}%
\bibitem [{\citenamefont {Wang}\ and\ \citenamefont
  {Lin}(2025)}]{Wang:2025mqz}%
  \BibitemOpen
  \bibfield  {author} {\bibinfo {author} {\bibfnamefont {Y.}~\bibnamefont
  {Wang}}\ and\ \bibinfo {author} {\bibfnamefont {W.}~\bibnamefont {Lin}},\
  }\href {\doibase 10.3847/1538-4357/adf336} {\bibfield  {journal} {\bibinfo
  {journal} {Astrophys. J.}\ }\textbf {\bibinfo {volume} {989}},\ \bibinfo
  {pages} {120} (\bibinfo {year} {2025})},\ \Eprint
  {http://arxiv.org/abs/2506.04333} {arXiv:2506.04333 [astro-ph.CO]}
  \BibitemShut {NoStop}%
\bibitem [{\citenamefont {Ibe}\ \emph {et~al.}(2019)\citenamefont {Ibe},
  \citenamefont {Yamazaki},\ and\ \citenamefont {Yanagida}}]{Ibe:2018ffn}%
  \BibitemOpen
  \bibfield  {author} {\bibinfo {author} {\bibfnamefont {M.}~\bibnamefont
  {Ibe}}, \bibinfo {author} {\bibfnamefont {M.}~\bibnamefont {Yamazaki}}, \
  and\ \bibinfo {author} {\bibfnamefont {T.~T.}\ \bibnamefont {Yanagida}},\
  }\href {\doibase 10.1088/1361-6382/ab5197} {\bibfield  {journal} {\bibinfo
  {journal} {Class. Quant. Grav.}\ }\textbf {\bibinfo {volume} {36}},\ \bibinfo
  {pages} {235020} (\bibinfo {year} {2019})},\ \Eprint
  {http://arxiv.org/abs/1811.04664} {arXiv:1811.04664 [hep-th]} \BibitemShut
  {NoStop}%
\bibitem [{\citenamefont {Lin}\ \emph {et~al.}(2023)\citenamefont {Lin},
  \citenamefont {Yanagida},\ and\ \citenamefont {Yokozaki}}]{Lin:2022khg}%
  \BibitemOpen
  \bibfield  {author} {\bibinfo {author} {\bibfnamefont {W.}~\bibnamefont
  {Lin}}, \bibinfo {author} {\bibfnamefont {T.~T.}\ \bibnamefont {Yanagida}}, \
  and\ \bibinfo {author} {\bibfnamefont {N.}~\bibnamefont {Yokozaki}},\ }\href
  {\doibase 10.1088/1572-9494/acb3b5} {\bibfield  {journal} {\bibinfo
  {journal} {Commun. Theor. Phys.}\ }\textbf {\bibinfo {volume} {75}},\
  \bibinfo {pages} {035203} (\bibinfo {year} {2023})},\ \Eprint
  {http://arxiv.org/abs/2209.12281} {arXiv:2209.12281 [hep-ph]} \BibitemShut
  {NoStop}%
\bibitem [{\citenamefont {{Foreman-Mackey}}\ \emph {et~al.}(2013)\citenamefont
  {{Foreman-Mackey}}, \citenamefont {{Hogg}}, \citenamefont {{Lang}},\ and\
  \citenamefont {{Goodman}}}]{emcee}%
  \BibitemOpen
  \bibfield  {author} {\bibinfo {author} {\bibfnamefont {D.}~\bibnamefont
  {{Foreman-Mackey}}}, \bibinfo {author} {\bibfnamefont {D.~W.}\ \bibnamefont
  {{Hogg}}}, \bibinfo {author} {\bibfnamefont {D.}~\bibnamefont {{Lang}}}, \
  and\ \bibinfo {author} {\bibfnamefont {J.}~\bibnamefont {{Goodman}}},\ }\href
  {\doibase 10.1086/670067} {\bibfield  {journal} {\bibinfo  {journal} {PASP}\
  }\textbf {\bibinfo {volume} {125}},\ \bibinfo {pages} {306} (\bibinfo {year}
  {2013})},\ \Eprint {http://arxiv.org/abs/1202.3665} {1202.3665} \BibitemShut
  {NoStop}%
\bibitem [{\citenamefont {Bhattacharya}\ \emph {et~al.}(2025)\citenamefont
  {Bhattacharya}, \citenamefont {Borghetto}, \citenamefont {Malhotra},
  \citenamefont {Parameswaran}, \citenamefont {Tasinato},\ and\ \citenamefont
  {Zavala}}]{Bhattacharya:2024kxp}%
  \BibitemOpen
  \bibfield  {author} {\bibinfo {author} {\bibfnamefont {S.}~\bibnamefont
  {Bhattacharya}}, \bibinfo {author} {\bibfnamefont {G.}~\bibnamefont
  {Borghetto}}, \bibinfo {author} {\bibfnamefont {A.}~\bibnamefont {Malhotra}},
  \bibinfo {author} {\bibfnamefont {S.}~\bibnamefont {Parameswaran}}, \bibinfo
  {author} {\bibfnamefont {G.}~\bibnamefont {Tasinato}}, \ and\ \bibinfo
  {author} {\bibfnamefont {I.}~\bibnamefont {Zavala}},\ }\href {\doibase
  10.1088/1475-7516/2025/04/086} {\bibfield  {journal} {\bibinfo  {journal}
  {JCAP}\ }\textbf {\bibinfo {volume} {04}},\ \bibinfo {pages} {086} (\bibinfo
  {year} {2025})},\ \Eprint {http://arxiv.org/abs/2410.21243} {arXiv:2410.21243
  [astro-ph.CO]} \BibitemShut {NoStop}%
\bibitem [{\citenamefont {Wolf}\ and\ \citenamefont
  {Ferreira}(2023)}]{Wolf:2023uno}%
  \BibitemOpen
  \bibfield  {author} {\bibinfo {author} {\bibfnamefont {W.~J.}\ \bibnamefont
  {Wolf}}\ and\ \bibinfo {author} {\bibfnamefont {P.~G.}\ \bibnamefont
  {Ferreira}},\ }\href {\doibase 10.1103/PhysRevD.108.103519} {\bibfield
  {journal} {\bibinfo  {journal} {Phys. Rev. D}\ }\textbf {\bibinfo {volume}
  {108}},\ \bibinfo {pages} {103519} (\bibinfo {year} {2023})},\ \Eprint
  {http://arxiv.org/abs/2310.07482} {arXiv:2310.07482 [astro-ph.CO]}
  \BibitemShut {NoStop}%
\bibitem [{\citenamefont {Wolf}\ \emph {et~al.}(2024)\citenamefont {Wolf},
  \citenamefont {Garc{\'\i}a-Garc{\'\i}a}, \citenamefont {Bartlett},\ and\
  \citenamefont {Ferreira}}]{Wolf:2024eph}%
  \BibitemOpen
  \bibfield  {author} {\bibinfo {author} {\bibfnamefont {W.~J.}\ \bibnamefont
  {Wolf}}, \bibinfo {author} {\bibfnamefont {C.}~\bibnamefont
  {Garc{\'\i}a-Garc{\'\i}a}}, \bibinfo {author} {\bibfnamefont {D.~J.}\
  \bibnamefont {Bartlett}}, \ and\ \bibinfo {author} {\bibfnamefont {P.~G.}\
  \bibnamefont {Ferreira}},\ }\href {\doibase 10.1103/PhysRevD.110.083528}
  {\bibfield  {journal} {\bibinfo  {journal} {Phys. Rev. D}\ }\textbf {\bibinfo
  {volume} {110}},\ \bibinfo {pages} {083528} (\bibinfo {year} {2024})},\
  \Eprint {http://arxiv.org/abs/2408.17318} {arXiv:2408.17318 [astro-ph.CO]}
  \BibitemShut {NoStop}%
\bibitem [{\citenamefont {Arkani-Hamed}\ \emph {et~al.}(2007)\citenamefont
  {Arkani-Hamed}, \citenamefont {Motl}, \citenamefont {Nicolis},\ and\
  \citenamefont {Vafa}}]{Arkani-Hamed:2006emk}%
  \BibitemOpen
  \bibfield  {author} {\bibinfo {author} {\bibfnamefont {N.}~\bibnamefont
  {Arkani-Hamed}}, \bibinfo {author} {\bibfnamefont {L.}~\bibnamefont {Motl}},
  \bibinfo {author} {\bibfnamefont {A.}~\bibnamefont {Nicolis}}, \ and\
  \bibinfo {author} {\bibfnamefont {C.}~\bibnamefont {Vafa}},\ }\href {\doibase
  10.1088/1126-6708/2007/06/060} {\bibfield  {journal} {\bibinfo  {journal}
  {JHEP}\ }\textbf {\bibinfo {volume} {06}},\ \bibinfo {pages} {060} (\bibinfo
  {year} {2007})},\ \Eprint {http://arxiv.org/abs/hep-th/0601001}
  {arXiv:hep-th/0601001} \BibitemShut {NoStop}%
\bibitem [{\citenamefont {Li}\ \emph {et~al.}(2025)\citenamefont {Li},
  \citenamefont {Wang}, \citenamefont {Zhang}, \citenamefont {Saridakis},\ and\
  \citenamefont {Cai}}]{Li:2025cxn}%
  \BibitemOpen
  \bibfield  {author} {\bibinfo {author} {\bibfnamefont {C.}~\bibnamefont
  {Li}}, \bibinfo {author} {\bibfnamefont {J.}~\bibnamefont {Wang}}, \bibinfo
  {author} {\bibfnamefont {D.}~\bibnamefont {Zhang}}, \bibinfo {author}
  {\bibfnamefont {E.~N.}\ \bibnamefont {Saridakis}}, \ and\ \bibinfo {author}
  {\bibfnamefont {Y.-F.}\ \bibnamefont {Cai}},\ }\href {\doibase
  10.1088/1475-7516/2025/08/041} {\bibfield  {journal} {\bibinfo  {journal}
  {JCAP}\ }\textbf {\bibinfo {volume} {08}},\ \bibinfo {pages} {041} (\bibinfo
  {year} {2025})},\ \Eprint {http://arxiv.org/abs/2504.07791} {arXiv:2504.07791
  [astro-ph.CO]} \BibitemShut {NoStop}%
\bibitem [{\citenamefont {Carroll}\ \emph {et~al.}(1990)\citenamefont
  {Carroll}, \citenamefont {Field},\ and\ \citenamefont
  {Jackiw}}]{Carroll:1989vb}%
  \BibitemOpen
  \bibfield  {author} {\bibinfo {author} {\bibfnamefont {S.~M.}\ \bibnamefont
  {Carroll}}, \bibinfo {author} {\bibfnamefont {G.~B.}\ \bibnamefont {Field}},
  \ and\ \bibinfo {author} {\bibfnamefont {R.}~\bibnamefont {Jackiw}},\ }\href
  {\doibase 10.1103/PhysRevD.41.1231} {\bibfield  {journal} {\bibinfo
  {journal} {Phys. Rev. D}\ }\textbf {\bibinfo {volume} {41}},\ \bibinfo
  {pages} {1231} (\bibinfo {year} {1990})}\BibitemShut {NoStop}%
\bibitem [{\citenamefont {Carroll}\ and\ \citenamefont
  {Field}(1991)}]{Carroll:1991zs}%
  \BibitemOpen
  \bibfield  {author} {\bibinfo {author} {\bibfnamefont {S.~M.}\ \bibnamefont
  {Carroll}}\ and\ \bibinfo {author} {\bibfnamefont {G.~B.}\ \bibnamefont
  {Field}},\ }\href {\doibase 10.1103/PhysRevD.43.3789} {\bibfield  {journal}
  {\bibinfo  {journal} {Phys. Rev. D}\ }\textbf {\bibinfo {volume} {43}},\
  \bibinfo {pages} {3789} (\bibinfo {year} {1991})}\BibitemShut {NoStop}%
\bibitem [{\citenamefont {Harari}\ and\ \citenamefont
  {Sikivie}(1992)}]{Harari:1992ea}%
  \BibitemOpen
  \bibfield  {author} {\bibinfo {author} {\bibfnamefont {D.}~\bibnamefont
  {Harari}}\ and\ \bibinfo {author} {\bibfnamefont {P.}~\bibnamefont
  {Sikivie}},\ }\href {\doibase 10.1016/0370-2693(92)91363-E} {\bibfield
  {journal} {\bibinfo  {journal} {Phys. Lett. B}\ }\textbf {\bibinfo {volume}
  {289}},\ \bibinfo {pages} {67} (\bibinfo {year} {1992})}\BibitemShut
  {NoStop}%
\bibitem [{\citenamefont {Workman}\ \emph {et~al.}(2022)\citenamefont {Workman}
  \emph {et~al.}}]{ParticleDataGroup:2022pth}%
  \BibitemOpen
  \bibfield  {author} {\bibinfo {author} {\bibfnamefont {R.~L.}\ \bibnamefont
  {Workman}} \emph {et~al.} (\bibinfo {collaboration} {Particle Data Group}),\
  }\href {\doibase 10.1093/ptep/ptac097} {\bibfield  {journal} {\bibinfo
  {journal} {PTEP}\ }\textbf {\bibinfo {volume} {2022}},\ \bibinfo {pages}
  {083C01} (\bibinfo {year} {2022})}\BibitemShut {NoStop}%
\bibitem [{\citenamefont {Ballardini}\ \emph {et~al.}(2025)\citenamefont
  {Ballardini}, \citenamefont {Gruppuso}, \citenamefont {Paradiso},
  \citenamefont {Sirletti},\ and\ \citenamefont {Natoli}}]{Ballardini:2025apf}%
  \BibitemOpen
  \bibfield  {author} {\bibinfo {author} {\bibfnamefont {M.}~\bibnamefont
  {Ballardini}}, \bibinfo {author} {\bibfnamefont {A.}~\bibnamefont
  {Gruppuso}}, \bibinfo {author} {\bibfnamefont {S.}~\bibnamefont {Paradiso}},
  \bibinfo {author} {\bibfnamefont {S.~S.}\ \bibnamefont {Sirletti}}, \ and\
  \bibinfo {author} {\bibfnamefont {P.}~\bibnamefont {Natoli}},\ }\href@noop {}
  {\  (\bibinfo {year} {2025})},\ \Eprint {http://arxiv.org/abs/2507.16714}
  {arXiv:2507.16714 [astro-ph.CO]} \BibitemShut {NoStop}%
\bibitem [{\citenamefont {Kallosh}\ \emph {et~al.}(1995)\citenamefont
  {Kallosh}, \citenamefont {Linde}, \citenamefont {Linde},\ and\ \citenamefont
  {Susskind}}]{Kallosh:1995hi}%
  \BibitemOpen
  \bibfield  {author} {\bibinfo {author} {\bibfnamefont {R.}~\bibnamefont
  {Kallosh}}, \bibinfo {author} {\bibfnamefont {A.~D.}\ \bibnamefont {Linde}},
  \bibinfo {author} {\bibfnamefont {D.~A.}\ \bibnamefont {Linde}}, \ and\
  \bibinfo {author} {\bibfnamefont {L.}~\bibnamefont {Susskind}},\ }\href
  {\doibase 10.1103/PhysRevD.52.912} {\bibfield  {journal} {\bibinfo  {journal}
  {Phys. Rev. D}\ }\textbf {\bibinfo {volume} {52}},\ \bibinfo {pages} {912}
  (\bibinfo {year} {1995})},\ \Eprint {http://arxiv.org/abs/hep-th/9502069}
  {arXiv:hep-th/9502069} \BibitemShut {NoStop}%
\bibitem [{\citenamefont {Berbig}(2025)}]{Berbig:2024aee}%
  \BibitemOpen
  \bibfield  {author} {\bibinfo {author} {\bibfnamefont {M.}~\bibnamefont
  {Berbig}},\ }\href {\doibase 10.1088/1475-7516/2025/03/015} {\bibfield
  {journal} {\bibinfo  {journal} {JCAP}\ }\textbf {\bibinfo {volume} {03}},\
  \bibinfo {pages} {015} (\bibinfo {year} {2025})},\ \Eprint
  {http://arxiv.org/abs/2412.07418} {arXiv:2412.07418 [astro-ph.CO]}
  \BibitemShut {NoStop}%
\end{thebibliography}%

\end{document}